\documentclass[3p,a4paper, 12pt]{elsarticle}
\pdfoutput=1

\usepackage{lineno, amsmath, amssymb,color,xcolor,xspace}
\usepackage{txfonts}
\usepackage{subfig}
\usepackage[normalem]{ulem}
\newcommand*{\eg}{{e.g.}\xspace}

\usepackage[normalem]{ulem}

\DeclareMathAlphabet{\pazocal}{OMS}{zplm}{m}{n}
\newcommand{\Q}{\pazocal{Q}}

\DeclareFontFamily{OT1}{pzc}{}
\DeclareFontShape{OT1}{pzc}{m}{it}{<-> s * [1.10] pzcmi7t}{}
\DeclareMathAlphabet{\mathpzc}{OT1}{pzc}{m}{it}
\newcommand{\ce}[1]{Eq.~(\ref{#1})}
\newcommand{\cf}[1]{{Fig.~\ref{#1}}}
\newcommand{\ct}[1]{{Table~\ref{#1}}}

\def\lapproxeq{\lower .7ex\hbox{$\;\stackrel{\textstyle                                                    <}{\sim}\;$}}                        \def\gapproxeq{\lower .7ex\hbox{$\;\stackrel{\textstyle                                                    >}{\sim}\;$}}

\newcommand{\msbar} {\overline{\text{MS}}}

\usepackage[skip=5pt plus1pt, indent=15pt]{parskip}

\usepackage{calc}
\journal{arXiv}

\DeclareMathAlphabet{\pazocal}{OMS}{zplm}{m}{n}
 \usepackage{ifpdf}
 \ifpdf
 \usepackage[pdftex]{hyperref}
 \else
 \usepackage[hypertex]{hyperref}
 \fi

 \hypersetup{
   pdftitle={Curing the high-energy perturbative instability of vector-quarkonium-photoproduction cross sections at order $\alpha \alpha_s^3$ with high-energy factorisation},%
   pdfauthor={},%
   pdfsubject={},%
   pdfkeywords={},%
   pdfstartview={},%
   bookmarksopen=true, breaklinks=true, debug=true, %
   colorlinks=true, linkcolor=red, citecolor=blue, urlcolor=blue
 }

\begin{document}

\begin{frontmatter}

\title{{Curing the high-energy perturbative instability of vector-quarkonium-photoproduction cross sections at order $\alpha \alpha_s^3$ with high-energy factorisation}}

%% or include affiliations in footnotes:

\author[a]{Jean-Philippe Lansberg}
\ead{Jean-Philippe.Lansberg@in2p3.fr}
\author[a,b]{Maxim Nefedov}
\ead{Maxim.Nefedov@ijclab.in2p3.fr}
\author[a]{Melih A. Ozcelik}
\ead{Melih.Ozcelik@ijclab.in2p3.fr}

\address[a]{Universit\'{e} Paris-Saclay, CNRS, IJCLab,  91405 Orsay, France}
\address[b]{National Centre for Nuclear Research (NCBJ), Pasteura 7, 02-093 Warsaw, Poland}

\begin{abstract}
{We cure the perturbative instability of the total-inclusive-photoproduction cross sections of vector $S$-wave quarkonia observed at high photon-proton-collision energies ($\sqrt{s_{\gamma p}}$)
in Next-to-Leading Order (NLO) Collinear-Factorisation (CF) computations. This is achieved
using  High-Energy Factorisation (HEF) in the {Doubly-Logarithmic Approximation (DLA)}, which is a subset of the {Leading-Logarithmic Approximation (LLA)} of HEF which resums higher-order {QCD} corrections proportional to $\alpha_s^n \ln^{n-1} (\hat{s}/M^2)$ in the Regge limit $\hat{s}\gg M^2$ with $M^2$ being the quarkonium mass and $\hat{s}$ is the squared partonic-center-of-mass energy. Such a DLA is strictly consistent with the NLO and NNLO DGLAP evolutions of the Parton Distribution Functions. By improving the treatment of the large-$\hat{s}$ asymptotics of the CF coefficient function, the resummation cures the unphysical results of the NLO CF calculation. The matching is directly performed in $\hat{s}$ space using the Inverse-Error Weighting matching procedure which avoids any possible double counting. 
The obtained cross sections are in good agreement with data. In addition, the scale-variation uncertainty of the matched result is significantly reduced compared to the LO results. Our calculations also yield closed-form analytic limits for $\hat{s}\gg M^2$ of the NLO partonic CF and numerical limits for contributions to those at NNLO scaling like $\alpha_s^2 \ln(\hat{s}/M^2)$.} 

\end{abstract}

\begin{keyword}
Heavy quarkonium, photoproduction, NLO perturbative calculations, Regge limit, high-energy factorisation,  resummation, matching
%\MSC[2010] 00-01\sep  99-00
\end{keyword}

\end{frontmatter}

%\linenumbers
{
\section{Introduction}
Historically the motivation for the study of the inclusive production of quarkonia in hadron-hadron and lepton-hadron collisions was to gain novel information on the structure of hadrons, see \eg \cite{Halzen:1984rq,Martin:1987ww}. Much experimental and theoretical effort has thus been devoted to it. We guide the reader to reviews~\cite{Lansberg:2019adr,Andronic:2015wma,Brambilla:2010cs,Lansberg:2006dh, Brambilla:2004wf,Kramer:2001hh} from which one quickly realises that, unfortunately, the leading mechanisms of inclusive-quarkonium-production reactions  generally remain unclear with several models of the non-perturbative hadronisation of quarkonia being used within the community. 

To achieve a better understanding of the non-perturbative dynamics of quarkonium production --whatever the motivation behind-- it is crucial to ensure some reliability of the perturbative part. Whereas one understands now that the quarkonium-transverse-momentum ($p_T$) distributions receive large radiative corrections which need to be properly dealt with\footnote{This is due to the more favourable $p_T$ scaling of a class of real-emission contributions~\cite{Artoisenet:2008fc,Lansberg:2008gk,Lansberg:2009db,Gong:2012ah,Lansberg:2013qka,Lansberg:2014swa,Lansberg:2017ozx,Shao:2018adj,Flore:2020jau}.}, it has been rediscovered a few years ago that $p_T$-integrated cross sections --referred to here as ``total" cross sections-- are plagued by perturbative instabilities~\cite{Lansberg:2010cn, Ozcelik:2019qze, Lansberg:2020ejc}.\footnote{This was initially noted in~\cite{Schuler:1994hy,Mangano:1996kg} and then simply forgotten.}

These were identified and cured within a strict NLO Collinear Factorisation (CF) set up by two of us in the case of pseudoscalar-quarkonium hadroproduction~\cite{Lansberg:2020ejc} and ($S$-wave) vector-quarkonium photoproduction~\cite{ColpaniSerri:2021bla}, which is the focus of the present paper. The observed negative and unphysical cross sections as well as the associated large observed factorisation-scale, $\mu_F$, dependence for both processes at high energies
were attributed to the subtraction of the collinear divergences into the Parton Distribution Functions (PDFs) in the $\msbar$ scheme. In the latter scheme, we could identify an over-subtraction of these divergences which then yield  negative partonic cross sections in regions where they ought to be positive in NLO computations. To cure this problem, we proposed~\cite{Lansberg:2020ejc} a new scale prescription, dubbed as $\hat \mu_F$, amounting to considering that NLO QCD radiative corrections in the {\it partonic} high-energy limit ($\hat{s} \to \infty$) should be accounted  for by the --positive definite\footnote{As we discussed in~\cite{Lansberg:2020ejc}, we believe that NLO gluon PDFs at low scales should be positive for quarkonium phenomenology to be well behaved in line with~\cite{Candido:2020yat}. However, see~\cite{Collins:2021vke} for arguments in favour of the possibility for negative PDFs.}-- PDFs,  as we recapitulate in Section~\ref{sec:CF}. 
In doing so, the resulting hadronic cross sections then became positive and, in the photoproduction case, relatively close to the data.

Having understood the origin of this instability, we wondered whether a theoretical setup {going beyond the NLO of CF} such as {{\it High-Energy Factorisation (HEF)}~\cite{Catani:1990xk,Catani:1990eg,Collins:1991ty,Catani:1994sq}, resumming the higher-order corrections to the CF coefficient function which are enhanced by $\ln(\hat{s}/M^2)$ at $\hat{s}\gg M^2$,} could address the problem in a more general manner. We thus performed a first study~\cite{Lansberg:2021vie} of the simpler case of pseudoscalar-quarkonium hadroproduction where the quarkonium is produced in a $2\to 1$ partonic process at LO in $\alpha_s$. 

HEF allows one to sum the series of higher-order corrections to the CF coefficient function proportional to  $\alpha_s^n \ln^{n-1}(\hat{s}/M^2)$ which, in the context of the present study, we refer to as the  {\it Leading-Logarithmic Approximation (LLA)}. Since the resummation of high-energy logarithms affects both the DGLAP evolution of the PDFs~\cite{DGLAP1,DGLAP2,DGLAP3} and the CF coefficient function, we have appropriately truncated the LLA resummation in the CF coefficient function down to the {\it Doubly-Logarithmic Approximation (DLA)}, as described in  Ref.~\cite{Lansberg:2021vie}. Doing so, one can use the standard fixed-order PDFs consistently with the resummed CF coefficient functions. 

On the other hand, the HEF formalism is valid only up to  $M^2/\hat{s}$-power-suppressed corrections. As such, it cannot provide a good approximation to the CF coefficient function where $M^2$ is not negligible with respect to $\hat{s}$. To avoid this shortcoming, we use a matching procedure to smoothly interpolate between the $\hat{s}\gg M^2$ asymptotics obtained from HEF and the NLO CF coefficient function at $\hat{s} \gtrsim M^2$. In Ref.~\cite{Lansberg:2021vie}, we have proposed a version of the {\it Inverse-Error-Weighting (InEW)} matching prescription, first introduced in Ref.~\cite{Echevarria:2018qyi}, which uses our perturbative knowledge to assess the weight-determination procedure of InEW. As a result, we have obtained perturbatively-stable predictions for the pseudoscalar-quarkonium hadroproduction cross sections with a scale-dependence reduced in comparison to LO CF predictions. 

While the case of pseudoscalar-quarkonium hadroproduction might be considered as academical owing to the obvious challenge to measure such a $p_T$-integrated cross section, experimental data exist for $J/\psi$ photoproduction~\cite{H1:1996kyo,Denby:1983az,NA14:1986mdd} and further studies could be performed. Let us cite data from AMBER~\cite{Adams:2018pwt} at CERN, those at the future EIC, even for $\Upsilon$~\cite{ColpaniSerri:2021bla}, and those at the LHC in ultra-peripheral collisions up to TeV photon-proton collision energies~\cite{TalkInclusiveOniumUPC,InclusiveOniumUPC}. In principle, inclusive $J/\psi$ photoproduction is an interesting source of information to constrain gluon PDFs at low scale $\mu_F$ and low $x$. This is why we study it here using HEF matched to CF to properly account for the entire energy region among possible future measurements.

The structure of the manuscript is as follows. In Section~\ref{sec:CF}, we review the computation of the total cross section of an $S$-wave vector-quarkonium photoproduction in CF. In Section~\ref{sec:HEF}, we explain how the HEF formalism is applied to this process and present the cross checks we have performed. In Section~\ref{sec:matching+num}, the InEW matching procedure is described. {Phenomenological predictions for $J/\psi$ and $\Upsilon(1S)$ photoproduction and a discussion of theoretical uncertainties are then presented in the same section.} Section~\ref{sec:conclusions} gathers our conclusion and an outlook on other quarkonium-production processes. \ref{AppendixA} presents the technical details of the computation of the $p_T$-integrated HEF coefficient function used in the present study and~\ref{AppendixB} provides the details  of the computation and results for the $\hat{s}\gg M^2$ asymptotics of NLO and of $\alpha_s^2\ln(\hat{s}/M^2)$ NNLO terms of the CF coefficient function.  }

\section{$S$-wave-vector-quarkonium photoproduction: collinear factorisation}
\label{sec:CF}

As announced, our focus will be on the process of inclusive photoproduction of a vector $S$-wave-quarkonium state, which we will denote $\Q$:
\begin{equation}
 \gamma(q) + p(P) \to \Q(p)+ X, \label{proc:gamma+p-jpsi+X}
\end{equation}
with the quarkonium mass {$M$ and} $p^2=M^2$ while $q^2=P^2=0$.

Assuming factorisation of the quarkonium-hadronisation process from the initial-state proton and its remnants, one can write down the CF formula for the total, $p_T$-integrated, cross section of the process of \ce{proc:gamma+p-jpsi+X}:
\begin{equation}
\sigma(\sqrt{s_{\gamma p}}, M, z_{\max})=\int\limits_{0}^{\eta_{\max}} \frac{d\eta}{\eta}\ \frac{d\sigma}{d\ln \eta}, 
\end{equation}
{with}
\begin{equation}
\frac{d\sigma}{d\ln\eta}=\sum\limits_{i=g,q,\bar{q}} \frac{d{\cal L}_i(\eta,\sqrt{s_{\gamma p}}, M,\mu_F)}{d\ln\eta} \hat{\sigma}_{\gamma i}(\eta,z_{\max},\mu_F,\mu_R), \label{eq:TCS-int-eta}
\end{equation} 
where $\hat{\sigma}_{\gamma i}$ is the CF partonic coefficient function for the partonic channel $\gamma(q)+i(p_1)\to \Q(p)+ X$. Note that we have chosen the dimensionless distance from the partonic threshold,
\begin{equation}
\eta=\frac{\hat{s}-M^2}{M^2},\label{eq:eta-def}
\end{equation} 
as our convolution variable in \ce{eq:TCS-int-eta}. As usual, $\hat{s}=(p_1+q)^2$ is the squared center-of-mass energy in the partonic subprocess. We also have that $\eta_{\max}=(s_{\gamma p}-M^2)/M^2$ in \ce{eq:TCS-int-eta}. The partonic luminosity factor in \ce{eq:TCS-int-eta} is defined as:
\begin{equation}
\frac{d{\cal L}_i(\eta,\sqrt{s_{\gamma p}}, M,\mu_F)}{d\ln \eta}=\frac{M^2\eta}{s_{\gamma p}}  f_i\left(\frac{M^2}{s_{\gamma p}}(1+\eta),\mu_F\right),\label{eq:L-def}
\end{equation}
 where $f_i(x,\mu_F)$ is the (number density) CF PDF for a parton of flavour $i=g,q,\bar{q}$ in the proton, whose factorisation-scale, $\mu_F$, dependence is governed by the DGLAP evolution equations. 
 
 Typically, in experimental measurements of $J/\psi$ photoproduction, one places a cut on the {\it elasticity} kinematical variable:
\begin{equation}
z= \frac{P\cdot p}{P\cdot q}, \label{eq:z-def}
\end{equation}
which represents the fraction of the large light-cone component of the photon momentum carried away by the vector meson. In the proton rest frame, it equivalently corresponds to the vector-meson energy divided by the photon energy. Indeed, one usually wishes to exclude from the inclusive data the large-$z$ region where exclusive production takes place and one imposes $z<z_{\max}<1$. The presence of this cut is indicated as dependence on $z_{\max}$ in  \ce{eq:TCS-int-eta} and below.

In the present paper, we will use the Colour-Singlet (CS)-dominance approximation\footnote{We stress that, in the present case, it amounts to simply consider NRQCD at leading order in $v^2$, Colour-Octet (CO) contributions being parametrically NNLO in $v^2$.} of the Non-Relativistic QCD (NRQCD) factorisation hypothesis~\cite{Bodwin:1994jh}, where the CF coefficient function for the quarkonium production is given by the product of the CF coefficient function for the production of a heavy quark-antiquark pair $Q\bar Q$ in the CS state with a total spin $S=1$ and a relative orbital momentum $L=0$ and the Long-Distance Matrix Element (LDME) describing the hadronisation of the $Q\bar Q$ pair to the observable quarkonium state. At LO in $\alpha_s$, only one partonic subprocess contributes to the coefficient function:
\begin{equation}
\gamma(q)+g(p_1) \to Q\bar{Q}\left[{}^3S_1^{[1]} \right](p) + g(k), \label{proc:gamma+g-cc+g}
\end{equation}
while, at NLO in $\alpha_s$, besides the virtual contributions via the interference of the one-loop and Born amplitudes of the subprocess \ce{proc:gamma+g-cc+g}, the following real-emission subprocesses also contribute:
\begin{eqnarray}
\gamma(q)+g(p_1) \to Q\bar{Q}\left[{}^3S_1^{[1]} \right](p) + g(k_1)+g(k_2), \label{proc:gamma+g-cc+gg} \\
\gamma(q)+g(p_1) \to Q\bar{Q}\left[{}^3S_1^{[1]} \right](p) + q(k_1)+\bar{q}(k_2), \label{proc:gamma+g-cc+qq} \\
\gamma(q)+q(p_1) \to Q\bar{Q}\left[{}^3S_1^{[1]} \right](p) + q(k_1)+g(k_2). \label{proc:gamma+q-cc+qg} 
\end{eqnarray}  

  These NLO contributions have been computed for the first time by Kr\"amer in 1995 ~\cite{Kramer:1995nb} and we have successfully reproduced~\cite{ColpaniSerri:2021bla} these results using the FDC code~\cite{Wang:2004du, Gong:2012ah} based on the phase-space slicing method~\cite{Harris-Owens-NLO} as well as by our in-house implementation of the NLO calculation, based on the dipole-subtraction method~\cite{Catani:1996vz}. The partonic coefficient function, which includes the process of \ce{proc:gamma+g-cc+g} at LO and the processes of Eqs.~(\ref{proc:gamma+g-cc+gg}-\ref{proc:gamma+q-cc+qg}) as well as the one-loop correction at NLO can be conveniently expressed as follows:
\begin{eqnarray}
  \hat{\sigma}^{\rm (CF)}_{\gamma g}(\eta, \mu_F,\mu_R, z_{\max})&=& F_{\rm LO} \left\{ c_0(\eta,z_{\max}) 
  + \frac{\alpha_s(\mu_R)}{2\pi} \left[ \beta_0(n_{l}) c_0(\eta,z_{\max}) \ln \frac{\mu_R^2}{\mu_F^2} \right.\right. \nonumber \\ &+& \left. \left. c_1^{(\gamma g)}(\eta, z_{\max}, n_l) + \bar{c}^{(\gamma g)}_1(\eta,z_{\max})\ln\frac{M^2}{\mu_F^2} \right]   \right\} , \label{eq:SFs-ga-g-def} \\
  \hat{\sigma}^{\rm (CF)}_{\gamma q}(\eta, \mu_F,\mu_R, z_{\max}) &=& F_{\rm LO} \frac{\alpha_s(\mu_R)}{2\pi} \left[ c_1^{(\gamma q)}(\eta,z_{\max}) + \bar{c}^{(\gamma q)}_1(\eta,z_{\max}) \ln\frac{M^2}{\mu_F^2} \right] , \label{eq:SFs-ga-q-def}
\end{eqnarray}
where 
\[F_{\rm LO}=\frac{16\alpha\alpha_s^2(\mu_R) e_Q^2}{9M^2} \frac{\langle {\cal O} \left[ {}^3S_1^{[1]}  \right]\rangle}{M^3},\]
with $\langle {\cal O} \left[ {}^3S_1^{[1]}  \right]\rangle$ being NRQCD LDME~\cite{Bodwin:1994jh} describing the transition of the CS $Q\bar{Q}\left[{}^{3}S_1^{[1]}\right]$ state to the observable quarkonium state $\Q$ and $e_Q$ the electric charge of the heavy quark in units of positron charge. At LO in $v^2$, the CS LDME is proportional to the squared potential-model radial wave function, $|R(0)|^2$, evaluated at the origin in the position space: $\langle {\cal O} \left[ {}^3S_1^{[1]}  \right]\rangle = 2C_A(2J+1) |R(0)|^2/(4\pi)$ with $J=1$ being the total spin of the produced hadron. The $n_l$ is the number of quark flavours $q$ lighter than the considered heavy-quark $Q$. The dimensionless scaling functions (SFs) $c_0(\eta,z_{\max})$, $c_1(\eta,z_{\max},n_l)$ and $\bar{c}_1(\eta,z_{\max})$ in \ce{eq:SFs-ga-g-def} and \ce{eq:SFs-ga-q-def} only depend on the partonic energy variable $\eta$ defined in \ce{eq:eta-def} and the kinematical cut on the variable $z$~(\ce{eq:z-def}). In addition, the SF $c_1^{(\gamma g)}$ depends on the number of light flavours $n_l$ due to the $q\bar{q}$ splitting in \ce{proc:gamma+g-cc+qq}. Our definition of these SFs differs from the original definition of Kr\"amer~\cite{Kramer:1995nb} as well as one of Ref.~\cite{ColpaniSerri:2021bla} by the usage of a more traditional expansion parameter, $\alpha_s/(2\pi)$, rather than $g_s^2$ { which has been used in Refs.~\cite{Kramer:1995nb,ColpaniSerri:2021bla},} as well as by the choice of the $\mu_F$-dependent logarithm $\ln(M^2/\mu_F^2)$ and the sign convention for $\bar{c}_1$. We stress that when summed together they yield the very same $\hat{\sigma}^{\rm (CF)}_{\gamma i}$.

\begin{figure}[hbt!]
\begin{center}
\begin{tabular}{cc}
  \includegraphics[width=0.545\textwidth]{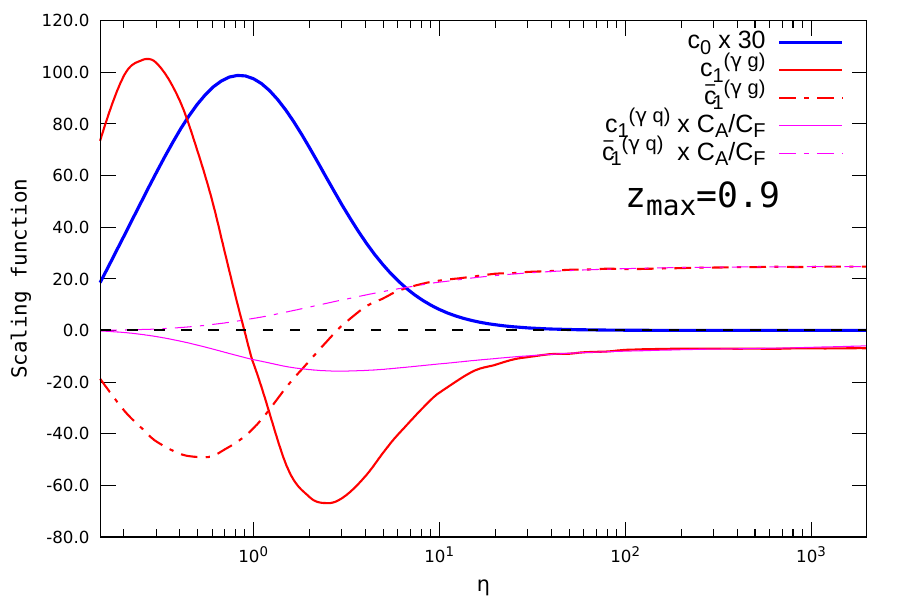} &  \includegraphics[width=0.35\textwidth]{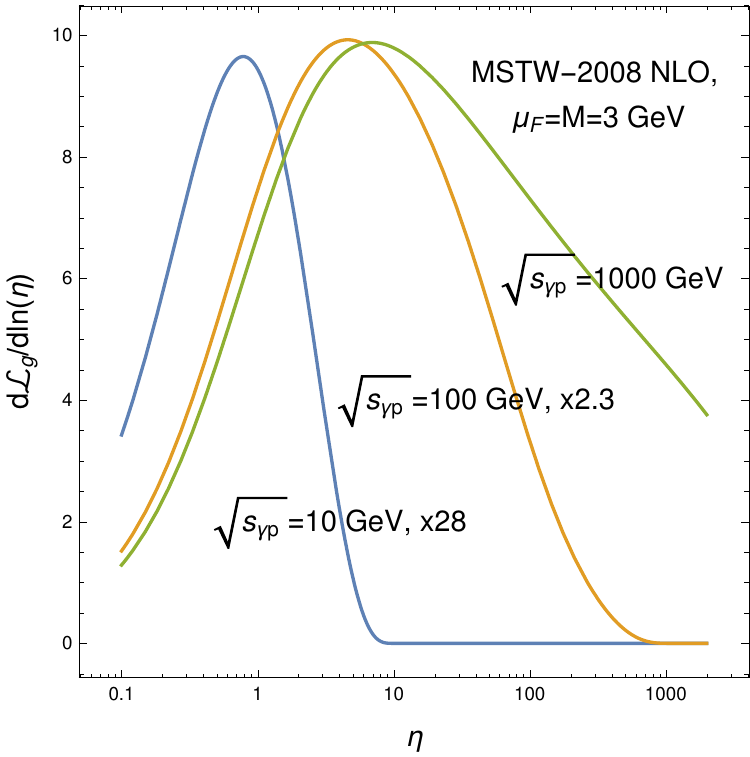} \\
  (a) & (b) 
\end{tabular}
\end{center}
\caption{Panel (a): Plots of the LO ($c_0$, thick solid blue line, multiplied by 30 for visibility) and NLO ($c_1^{(\gamma g)}$, solid red line; $c_1^{(\gamma q)}$, solid magenta line; $\bar{c}_1^{(\gamma g)}$, dash-dotted red line; $\bar{c}_1^{(\gamma q)}$,  dash-dotted magenta line. Note that the quark SFs are multiplied by $C_A/C_F$) scaling functions entering \ce{eq:SFs-ga-g-def} and \ce{eq:SFs-ga-q-def} as functions of $\eta$. For the plot, we have set $n_l=3$ for the coefficient function $c_1^{(\gamma g)}$. \label{fig:c0_c1_c1b-vs-eta} Panel (b): plots of the gluon luminosity factor \ce{eq:L-def} for $M=\mu_F=3$ GeV as a function of $\eta$ for several values of $\sqrt{s_{\gamma p}}$ using the MSTW 2008 NLO PDF central set~\cite{Martin:2009iq} for illustration.}
\end{figure}

The partonic-energy dependence of the SFs defined above is illustrated in  \cf{fig:c0_c1_c1b-vs-eta}(a).  The SF $c_0$ is decreasing as $1/\eta^2$, while all the NLO SFs tend to constant values in the high-energy limit. The asymptotic value of the SF $c^{(\gamma g)}_1(\eta,z_{\max}=0.9)$ turns out to be approximately $-6.978$ and that of $\bar{c}^{(\gamma g)}_1(\eta,z_{\max}=0.9)$ is $24.76$. In addition, the asymptotic values of the SFs of the $\gamma q$ channel are related to those in the $\gamma g$ channel via Casimir scaling, as shown in  \cf{fig:c0_c1_c1b-vs-eta}(a).

With increasing collision energies, $\sqrt{s_{\gamma p}}$, the partonic luminosity factor, \ce{eq:L-def}, evaluated at $\mu_F\sim M \sim 3$ GeV, no longer suppresses contributions of large values of $\eta$ to the integral \ce{eq:TCS-int-eta} (see the \cf{fig:c0_c1_c1b-vs-eta}(b)). Given the constant behaviour of the NLO corrections to $\hat{\sigma}_i$ {at large $\eta$}, the region {where} $\eta\gg 1$ increasingly contributes to the cross section at high $\sqrt{s_{\gamma p}}\gg M$. This leads~\cite{ColpaniSerri:2021bla} to large negative NLO corrections to the inclusive photoproduction cross section of $J/\psi$ since the asymptotic value of $c^{(\gamma i)}_1$ is negative and to a catastrophically strong $\mu_F$ dependence at $\sqrt{s_{\gamma p}}\gg M$. This signals the instability of the perturbative expansion of this observable due to missing large higher-order corrections at $\eta\gg 1$. All these features are clearly visible in the plot in the \cf{fig:TCS_Jpsi_Y1S_NLO}(a). 
Large negative NLO corrections are present even in the case of bottomonium production (\cf{fig:TCS_Jpsi_Y1S_NLO}(b)). The scale-variation band in \cf{fig:TCS_Jpsi_Y1S_NLO} and all the other total-cross-section plots below are obtained through the 5-point scale-variation procedure, i.e. as an envelope of the cross section curves with $\mu_F=\mu_0 \cdot 2^{\zeta_1}$ and $\mu_R=\mu_0 \cdot 2^{\zeta_2}$ taking $(\zeta_1,\zeta_2)\in \{ (0,0),(0,\pm 1), (\pm 1,0)\}$, where $\mu_0$ is the central-scale choice, e.g. $\mu_0=M$ in \cf{fig:TCS_Jpsi_Y1S_NLO}. To obtain the numerical value of the $\langle {\cal O} \left[ {}^3S_1^{[1]}  \right]\rangle$ LDME we use the same values of $|R(0)|^2$ as used in the study~\cite{ColpaniSerri:2021bla}:  $1.25$ GeV$^3$ for $J/\psi$ and $7.5$ GeV$^3$ for $\Upsilon(1S)$. The estimates of feed-down production from excited states are also included in our plots using the same method as in Ref.~\cite{ColpaniSerri:2021bla}.

\begin{figure}[hbt!]
\begin{center}
\begin{tabular}{cc}
 \includegraphics[width=0.47\textwidth]{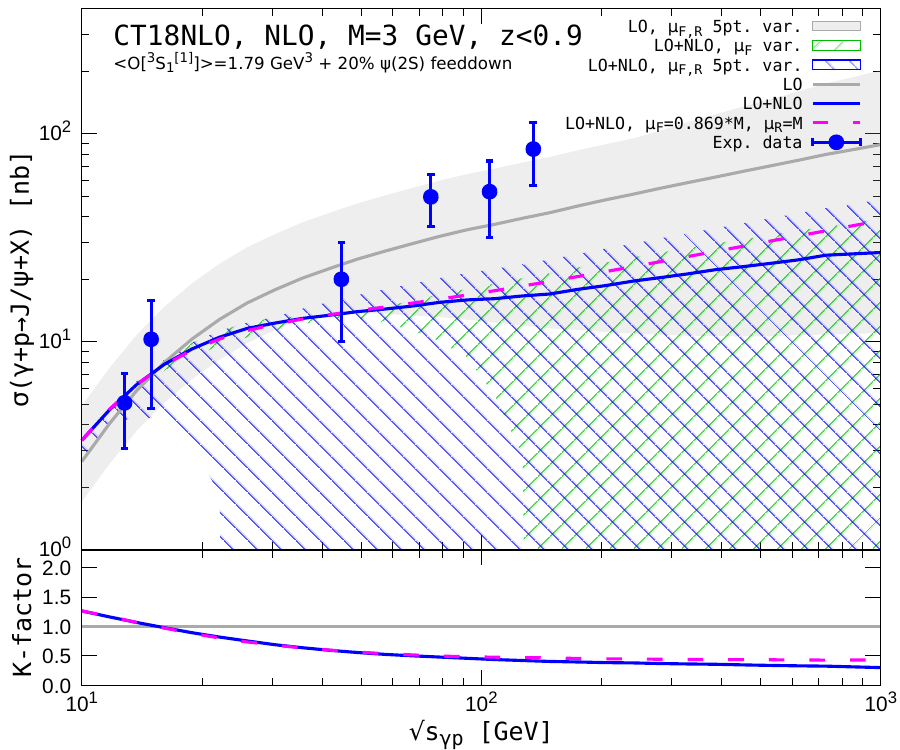} & \includegraphics[width=0.47\textwidth]{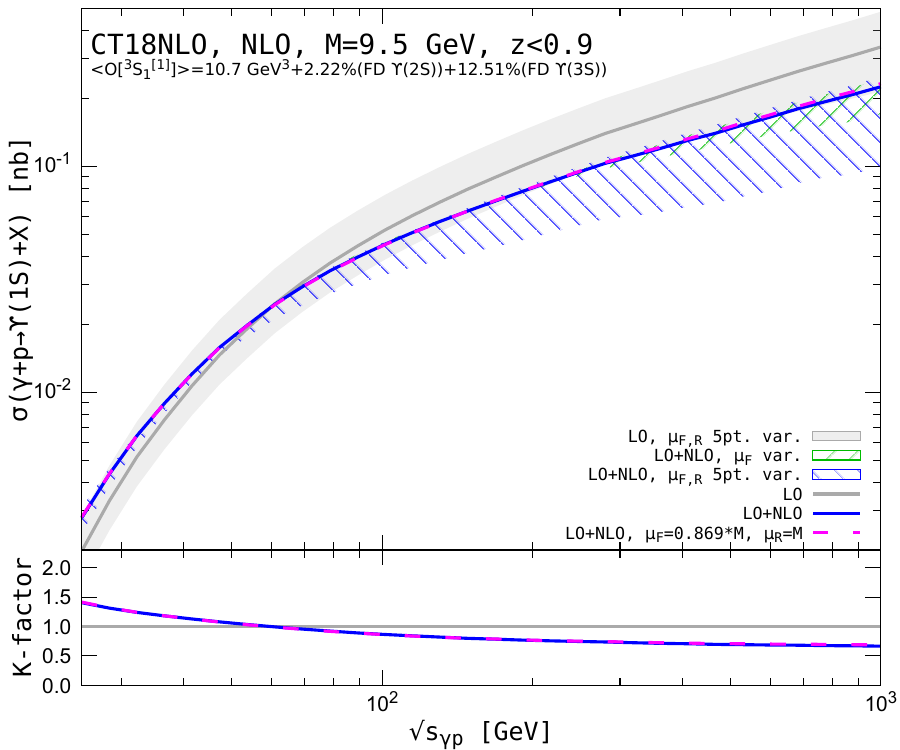}\\
 (a) & (b) 
\end{tabular}
\end{center}
\caption{Total inclusive photoproduction cross sections of $J/\psi$ (a) and $\Upsilon(1S)$ (b) in CF at LO and NLO in $\alpha_s$ and in NRQCD at LO in $v^2$ (CSM). The solid curves correspond to the default scale choice, $\mu_F=\mu_R=M$, while the shaded bands correspond to the 5-point $\mu_F$ and $\mu_R$ variation prescription described in the text. The dashed curve corresponds to the NLO computation with $\mu_F=\hat{\mu}_F$ and $\mu_R=M$. The experimental data in the left plot are taken from the H1~\cite{H1:1996kyo}, FTPS~\cite{Denby:1983az} and NA14~\cite{NA14:1986mdd} collaborations.\label{fig:TCS_Jpsi_Y1S_NLO}}
\end{figure}

As a matter of fact, similar perturbative instabilities of $p_T$-integrated cross sections at high collision energies have been observed in the 1990's in heavy-quarkonium physics as soon as first NLO computations appeared~\cite{Schuler:1994hy,Mangano:1996kg} and then were essentially forgotten. Their existence was restated in 2010~\cite{Lansberg:2010cn} and discussed in some details in full NRQCD in 2015 for $\eta_c$ and $J/\psi$~\cite{Feng:2015cba}. Only in 2020, were a first convincing diagnosis and a first solution with CF proposed for $\eta_c$ hadroproduction~\cite{Lansberg:2020ejc}.  The case of $J/\psi$ photoproduction was then discussed in 2021~\cite{ColpaniSerri:2021bla}. 

The solution proposed in Refs.~\cite{Lansberg:2020ejc,ColpaniSerri:2021bla} to cure these instabilities is based on the fact that the partonic high-energy limits of the scaling functions for all partonic channels are related by Casimir scaling (see the \cf{fig:c0_c1_c1b-vs-eta}(a)). As such, a unique factorisation-scale choice for all channels can make all scaling functions tend to zero at large $\eta$ and thence remove the large NLO corrections coming from the $\eta\gg 1$ region. It acts as if they were effectively absorbed  into the PDF evolution. Considering \ce{eq:SFs-ga-g-def} and \ce{eq:SFs-ga-q-def} at $\eta\to\infty$, one finds such an optimal-scale value for the photoproduction case:
\begin{equation}
 \hat{\mu}_F=M\exp\left[\frac{c^{(\gamma q/g)}_1(\eta\to\infty,z_{\max})}{2\bar{c}^{(\gamma q/g)}_1(\eta\to\infty,z_{\max})} \right],\label{eq:muF-hat-def}
\end{equation}
which evaluates to $\hat{\mu}_F\simeq 0.869 M$ for $z_{\max}=0.9$. Predictions corresponding to this scale choice are plotted in \cf{fig:TCS_Jpsi_Y1S_NLO} as well as in \cf{fig:TCS_plots_Jpsi_zmax-09_NO-DATA} with dashed lines.

The $\hat{\mu}_F$ prescription of \ce{eq:muF-hat-def} legitimately exploits the factorisation-scale ambiguity of the fixed-order CF calculation. One could {therefore} explore the question of whether the DGLAP evolution of the PDFs {\it alone} could correctly capture the high-energy structure of higher-order QCD corrections to the partonic cross section. In other words, if one reexpresses the NLO calculation, done with PDFs at the scale $\mu_F=\hat{\mu}_F$, in terms of PDFs at a different scale $\mu_F=M$, will the higher-order corrections (NNLO and beyond) arising from the perturbative expansion of the DGLAP evolution of PDFs between the scales $\hat{\mu}_F$ and $M$ reproduce the higher-order corrections to $\hat{\sigma}_i(\eta\gg 1, \mu_F=M,\mu_R,z_{\max})$ to be obtained in an actual N$^{k>1}$LO computation of this object? Unfortunately, the answer to this question is negative even in the LLA, i.e. when only considering in $\hat{\sigma}_i$ terms  which are proportional to $\alpha_s^n \ln^{n-1}(1+\eta)$ at $\eta\gg 1$. Indeed, the coefficients in front of these terms can not be correctly reproduced by the DGLAP evolution {\it alone}. Instead, a more complicated formalism like HEF is required to calculate and resum those terms. 
\section{High-energy factorisation for $S$-wave-vector-quarkonium photoproduction}
\label{sec:HEF}

\subsection{Basic factorisation formula and coefficient functions of HEF}
\label{sec:HEF-basic}
\begin{figure}
    \centering
    \includegraphics{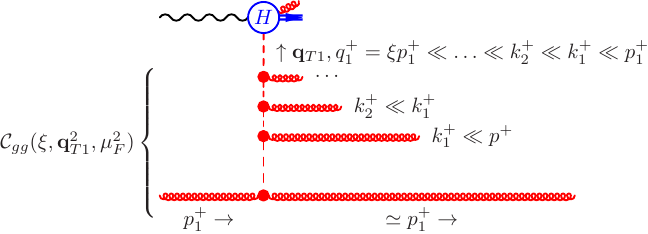}
    \caption{Typical Feynman diagram and Multi-Regge kinematics of emissions in the CF coefficient function of the process \ce{proc:ga+gq-cc+X} corresponding to the partonic high-energy ($\eta \gg 1$) limit in the LLA. The dashed line in the $t$ channel represents a {\it Reggeised gluon} and the solid circles represent Lipatov's vertices, while the open circle corresponds to the ``off-shell'' subprocess of \ce{proc:ga+R-3S11+g}. }
    \label{fig:HEF-photoprod}
\end{figure}

In the resummation part of our calculation, we are going to consider the general N$^{k\geq 1}$LO partonic subprocess:
\begin{equation}
\gamma(q)+ g/q(p_1)\to Q\bar{Q}\left[{}^3S_1^{[1]}\right](p) + X, \label{proc:ga+gq-cc+X}    
\end{equation}
with $q^2=0$ and $|{\bf q}_T|=0$ yielding to the following Sudakov components\footnote{We define our Sudakov decomposition in terms of the dimensionless vectors $n_-^\mu=P^\mu/P_+=(1,0,0,1)^\mu$ and $n_+^\mu=(1,0,0,-1)^\mu${, where the components are given in the $\gamma p$ center-of-mass frame,} such that $n_{\pm}^2=0$, $n_+n_-=2$. For any four-momentum $k^\mu=(k^+n_-^\mu + k^-n_+^\mu)/2 + k_T^\mu$ with $n_{\pm}\cdot k_T=0$, $k^{\pm}=n_{\pm}\cdot k$ and $k^2=k_+k_--{\bf k}_T^2$. The upper/lower position of indices $\pm$ does not have any meaning.} of the photon momentum moving in the negative z direction such that $q_+=0$ and $q_->0$. Moreover, the momentum of the incoming {\it collinear parton} $p^\mu_1=x_1P^\mu$ is such that $|{\bf p}_{1T}|=0$, $p_1^-=0$ and $p_1^+=x_1 P^+$. 

In the LLA at $\eta\gg 1$, the process \ce{proc:ga+gq-cc+X} is factorised into two stages, as depicted diagrammatically in  \cf{fig:HEF-photoprod}. First, the $k^+$-ordered cascade of real emissions carries away all but a tiny fraction of the initial $p_1^+$ momentum, $\xi$. At $\eta\gg 1$, $\xi$ will be of order $1/\eta$. This sequence of emissions leads to the appearance of large logarithms scaling like $\ln 1/\xi$ at each order in $\alpha_s$ and is described in the HEF formalism by the resummation factor ${\cal C}_{gg}(\xi, {\bf q}_{T1}^2,\mu^2_F,\mu^2_R)$ or ${\cal C}_{gq}(\xi,{\bf q}_{T1}^2,\mu^2_F,\mu^2_R)$, depending on whether the emission with largest $k^+$ in the cascade was a gluon or a quark. We stress that, in the LLA at leading power in ${\cal O}(1/\eta)$, all other emissions should be of gluons.  In the second stage, the $Q\bar{Q}$ pair is produced  via the fusion of a {\it Reggeised gluon} ($R_+$), moving in the proton direction and carrying the four momentum $q_1^{\mu} = \xi x_1 P^\mu+q_{T1}^\mu$ and of the photon:
\begin{equation}
    \gamma(q)+R_+(q_1)\to Q\bar{Q}\left[{}^3S_1^{[1]}\right](p) + g(k), \label{proc:ga+R-3S11+g}
\end{equation}
where the final-state gluon is necessary by virtue of colour and charge-parity conservation~\cite{Berger:1980ni}. The corresponding HEF coefficient function can be calculated using the following prescription:
\begin{equation}
    H(\bar{s},\bar{t},\bar{u},({\bf q}_{T1}\cdot {\bf p}_T), {\bf q}_{T1}^2)=\frac{(q_1^+)^2}{4 {\bf q}_{T1}^2} n_-^\mu n_-^\nu {\cal M}_{\mu\nu} = \frac{q_{T1}^\mu q_{T1}^\nu}{{\bf q}_{T1}^2} {\cal M}_{\mu\nu}, \label{eq:H-def}
\end{equation}
where the tensor ${\cal M}_{\mu\nu}$ is the squared  QCD amplitude of the process \ce{proc:gamma+g-cc+g}: 
\begin{enumerate}
    \item summed over the polarisations of the photon, quarkonium and final-state gluon, but not contracted with respect to the polarisation indices of the initial-state gluon in both the amplitude and its complex conjugate
    \item with the momentum of the initial on-shell gluon $p_1$ simply replaced by that of the off-shell Reggeised gluon, namely $q_1$. 
    \item accordingly, the Mandelstam variables for the subprocess \ce{proc:ga+R-3S11+g} are defined as $\bar{s}=(q+q_1)^2$, $\bar{t}=(q_1-p)^2$, $\bar{u}=(q-p)^2$. 
\end{enumerate}

The two equalities in \ce{eq:H-def} hold due to the Ward–Takahashi-like\footnote{As opposed to Slavnov–Taylor identities usually applicable in non-Abelian theories like QCD.} identities $q_1^\mu {\cal M}_{\mu\nu}=0$ and $q_1^\nu{\cal M}_{\mu\nu}=0$ which {are satisfied by the} tensor ${\cal M}_{\mu\nu}$ of the process \ce{proc:gamma+g-cc+g} even with $q_1^2<0$. The first equality in \ce{eq:H-def} uses the simplest coupling between a Reggeised gluon and a ``QCD'' gluon from the infinite tower of such couplings existing in Lipatov's EFT for multi-Regge processes in QCD~\cite{Lipatov95}. The second equality is referred to as the Gribov's trick  and shows that a smooth ${\bf q}_{T1}^2\to 0$ limit of \ce{eq:H-def} should exist.  

All what is needed to perform LLA computations  is the coefficient function $H$ at LO in $\alpha_s$ for the process \ce{proc:ga+R-3S11+g} first derived in 2004~\cite{Lipatov:2004ar}.  In our calculation, we will use a more compact expression~\cite{Kniehl:2006sk}: 

\begin{equation}
\begin{split}
&  H(\bar{s},\bar{t},\bar{u},{\bf q}_{T1}\cdot{\bf p}_T,{\bf q}_{T1}^2)=\pi^3\alpha \alpha_s^2
e_Q^2\frac{\langle{\cal O}[^3S_1^{[1]}]\rangle}{M^3}\,
\frac{2048 M^2}{27 (M^2 - \bar{s})^2 (M^2 - \bar{u})^2 ({\bf q}_{T1}^2 +
M^2 - \bar{t})^2}
\\
&\times \Bigl[ ({\bf q}_{T1}^2)^4 M^2 + M^2 \bigl(\bar{s}^2 + \bar{s}
\bar{u} + \bar{u}^2 - M^2 (\bar{s} + \bar{u})\bigr)^2 + ({\bf q}_{T1}^2)^3
(M^2 (5 \bar{s} + 3 \bar{u})-7 M^4 - \bar{s} \bar{u})
\\
&+ ({\bf q}_{T1}^2)^2 (\bar{s} \bar{u} (\bar{u}-\bar{s})+ M^4 (3 \bar{u}-11 \bar{s}) + M^2 (7 \bar{s}^2 + 2 \bar{s} \bar{u} - 3
\bar{u}^2)) + {\bf q}_{T1}^2 \bar{s} (\bar{s} \bar{u}^2 + M^4 (\bar{u}-6 \bar{s})
\\
& + M^2 (4 \bar{s}^2 + \bar{s} \bar{u} - \bar{u}^2))- 2
({\bf q}_{T1}\cdot{\bf p}_T) (({\bf q}_{T1}^2)^3 M^2 + ({\bf q}_{T1}^2)^2 (-7 M^4 - \bar{s}
\bar{u} + M^2 (3 \bar{s} + 4 \bar{u}))
\\
& + {\bf q}_{T1}^2 (M^4 (-7 \bar{s} + 2 \bar{u}) -\bar{s}^2 \bar{u}+
M^2 (2 \bar{s}^2 + \bar{s} \bar{u} - 2 \bar{u}^2)) - M^2 (2
M^4 (\bar{s} + \bar{u}) - 2 M^2 \bar{u} (3 \bar{s} + 2 {
\bar{u}})
\\
& + \bar{u} (3 \bar{s}^2 + 4 \bar{s} {
\bar{u}} + 2 \bar{u}^2))) - 2 M^2 \frac{({\bf q}_{T1}\cdot{\bf p}_T)^2}{{\bf q}_{T1}^2} (({\bf q}_{T1}^2)^3 + M^2 \bar{s}^2 + ({\bf q}_{T1}^2)^2 (M^2 + 2 \bar{s})
\\
&+ {\bf q}_{T1}^2 (2 M^2 \bar{s} + \bar{s}^2 - 2 \bar{t}^2)) \Bigr],
\end{split}
\label{eq:H_gamma+R-cc+g}
\end{equation}
where $e_Q$ is the electric charge of the heavy quark in units of the positron charge. The HEF coefficient function satisfies the following on-shell-limit property:
\begin{equation}
    \int\limits_0^{2\pi} \frac{d\phi}{2\pi} \lim\limits_{{\bf q}_{T1}^2\to 0} H(\bar{s},\bar{t},\bar{u},{\bf q}_{T1}\cdot {\bf p}_T, {\bf q}_{T1}^2) = \frac{(-g^{\mu\nu})}{2}{\cal M}_{\mu\nu}, \label{eq:H-HEF_on-shell}
\end{equation}
where $\phi$ is the azimuthal angle of ${\bf q}_{T1}$ relative to ${\bf p}_T$, which remains on the l.h.s. even after taking the limit ${\bf q}_{T1}^2\to 0$. On the r.h.s. of \ce{eq:H-HEF_on-shell}, the usual on-shell kinematics ($q_1^2=0$) for the tensor ${\cal M}_{\mu\nu}$ of the process \ce{proc:gamma+g-cc+g} is to be used as ${\bf q}_{T1}^2\to 0$. 

As  mentioned before, HEF allows one to resum a class of logarithmically-enhanced QCD corrections to CF. As such, the HEF partonic coefficient function can be used as a CF partonic coefficient function and convoluted with usual PDFs but this should in principle only be done in the region where the high-energy leading-power approximation and the LLA are applicable, i.e. $\eta \to \infty$. The general HEF formula for the CF partonic coefficient function for the process \ce{proc:ga+gq-cc+X} has the form~\cite{Catani:1990xk,Catani:1990eg,Collins:1991ty,Catani:1994sq}:
\begin{eqnarray}
\frac{d\hat{\sigma}^{\rm (HEF)}_{\gamma i}}{d\Pi_f}&=& \int\limits_{\xi_{\min}}^1 \frac{d\xi}{\xi} \int\frac{d^2{\bf q}_{T1}}{\pi} {\cal C}_{gi}(\xi,{\bf q}_{T1}^2,\mu_F^2,\mu_R^2) \nonumber \\ &\times& \frac{H(\bar{s},\bar{t},\bar{u},({\bf q}_{T1}\cdot {\bf p}_T), {\bf q}_{T1}^2)}{2M^2\xi(1+\eta)} \delta^{(4)}(q+q_1-p-k), 
\end{eqnarray}
where $\xi_{\min}=1/z/(1+\eta)$, so that $z\geq 1/(1+\eta)$. $d\Pi_f$ is the usual Lorentz-invariant phase-space volume element for the final-state particles with momenta $p$ and $k$. Integrating out the momentum-conserving $\delta$ functions, one can derive the following master formula for the $z$ and $p_T$-differential partonic cross section:
\begin{equation}
   \frac{d\hat{\sigma}^{\text{(HEF, $\ln 1/\xi$)}}_{\gamma i}}{dz d^2{\bf p}_T} =  \frac{1}{2M^2}\int\limits \frac{d^2{\bf q}_{T1}}{\pi} \int\limits_{\xi_{\min}}^{1} \frac{d\xi}{\xi}\ {\cal C}_{gi}\left(\xi,{\bf q}_{T1}^2,\mu_F^2,\mu_R^2 \right) \frac{d{\cal H}\big(\xi(1+\eta),z,{\bf q}_{T1},{\bf p}_T\big)}{dz d^2{\bf p}_T}, \label{eq:master-sig-HEF}
\end{equation}
where 
\begin{equation}
\begin{split}\frac{d{\cal H}(y,z,{\bf q}_{T1},{\bf p}_T)}{dz d^2{\bf p}_T} &= \frac{H(\bar{s},\bar{t},\bar{u},({\bf q}_{T1}\cdot {\bf p}_T), {\bf q}_{T1}^2)}{2(2\pi)^2 yz} \\ &\times \delta\left( (1-z) \left( M^2 y - \frac{M^2+{\bf p}_T^2}{z} \right) - ({\bf q}_{T1}-{\bf p}_T)^2  \right),\end{split}\label{eq:H-cal-def}
\end{equation}
where $y=\xi(1+\eta)$ and Mandelstam variables can be expressed in terms of $y$, $z$, ${\bf q}_{T1}^2$ and ${\bf p}_T^2$ as:
\begin{equation}
\begin{split}
\bar{s}&=M^2 y - {\bf q}_{T1}^2, \\
\bar{t}&=-\frac{1}{z}\left[ M^2 (yz-1)-{\bf p}_T^2 \right], \\
\bar{u}&= -\frac{1}{z}\left[ M^2(1-z)+{\bf p}_T^2 \right].
\end{split}
\label{eq:stu-sub}
\end{equation}
Due to the remaining $\delta$ function in \ce{eq:H-cal-def}, one has the identity:
\begin{equation}
    {\bf q}_{T1}\cdot {\bf p}_T=\frac{1}{2z}\left[ {\bf p}_T^2 + z{\bf q}_{T1}^2 -M^2(z\xi(1+\eta)-1)(1-z) \right], \label{eq:qT1.pT-sub}
\end{equation}
which, together with \ce{eq:stu-sub}, allows one to remove all explicit scalar products of the transverse momenta from \ce{eq:H_gamma+R-cc+g} and express these purely as functions of ${\bf p}_T^2$, ${\bf q}_{T1}^2$, $\eta$, $z$ and $\xi$. 

To compute the HEF contribution to the total $p_T$-integrated-quarkonium-photoproduction cross section, one simply has to integrate \ce{eq:master-sig-HEF} and thus only \ce{eq:H-cal-def} over ${\bf p}_T$ as it appears nowhere else. It turns out that the integration over ${\bf p}_T$ can be carried out in a closed analytic form at fixed $z$, which is very useful for numerical calculations. We explain the integration technique and give the corresponding analytic result for the ${\bf p}_T$-integrated function \ce{eq:H-cal-def} in \ref{AppendixA}.

\subsection{Strict LLA in $\ln(1+\eta)$\label{sec:strict-LLA}}

The resummation of $\ln(1/\xi)$ large logarithms to all orders in $\alpha_s$ in \ce{eq:master-sig-HEF} is provided by the resummation functions ${\cal C}_{gi}$ which, in the LLA in $\ln(1/\xi)$, have the following expansions in powers of $\alpha_s\ln\frac{1}{\xi}$:
\begin{equation}
    {\cal C}_{gi}(\xi,{\bf q}_T^2,\mu_F^2,\mu_R^2) = \sum\limits_{n=0}^\infty  \left[\alpha_s(\mu_R)\ln\frac{1}{\xi}\right]^n {\cal C}_{gi}^{(n)}({\bf q}_T^2,\mu_F^2).\label{eq:C-ser}
\end{equation}
Substituting this expansion into \ce{eq:master-sig-HEF} and making the change of integration variable $\xi=y/(1+\eta)$, one obtains:
\begin{eqnarray}
&&  \begin{split} \frac{d\hat{\sigma}^{\text{(HEF, $\ln 1/\xi$)}}_{\gamma i}}{dz d^2{\bf p}_T} &=  \frac{1}{2M^2} \int\limits \frac{d^2{\bf q}_{T1}}{\pi}  \sum\limits_{n=0}^\infty {\cal C}^{(n)}_{gi}\left({\bf q}_{T1}^2,\mu_F^2\right) \\ & \times\int\limits_{1/z}^{1+\eta} \frac{dy}{y} \left[ \alpha_s(\mu_R) \left( \ln(1+\eta) - \ln y  \right) \right]^n \frac{d{\cal H}\big(y,z,{\bf q}_{T1},{\bf p}_T\big)}{dz d^2{\bf p}_T} \end{split} \nonumber \\
&&   =  \frac{1}{2M^2} \int\limits \frac{d^2{\bf q}_{T1}}{\pi}  \sum\limits_{n=0}^\infty {\cal C}^{(n)}_{gi}\left({\bf q}_{T1}^2,\mu_F^2\right) \left[ \alpha_s(\mu_R) \ln(1+\eta)  \right]^n \int\limits_{1/z}^{1+\eta} \frac{dy}{y} \frac{d{\cal H}\big(y,z,{\bf q}_{T1},{\bf p}_T\big)}{dz d^2{\bf p}_T} \label{eq:sig-HEF-log-exp} \\
&& \begin{split} -  (n+1)\frac{\alpha_s(\mu_R)}{2M^2} \int\limits \frac{d^2{\bf q}_{T1}}{\pi}  \sum\limits_{n=0}^\infty {\cal C}^{(n+1)}_{gi}\left({\bf q}_{T1}^2,\mu_F^2\right) \left[ \alpha_s(\mu_R) \ln(1+\eta)  \right]^n \\ \times \int\limits_{1/z}^{1+\eta} \frac{dy}{y} \ln y \frac{d{\cal H}\big(y,z,{\bf q}_{T1},{\bf p}_T\big)}{dz d^2{\bf p}_T} +\ldots , \end{split} \nonumber
\end{eqnarray}
where additional terms arising from the expansion of $\left( \ln(1+\eta) - \ln y  \right)^n$ are denoted by the ellipsis.

If the function ${\cal H}$ decreases like a power law at large $y\sim (1+\eta)$, which is true for the case at hand, then the integrals over $y$ converge and therefore only the first term in \ce{eq:sig-HEF-log-exp} {is an} expansion in $[\alpha_s \ln(1+\eta)]^n$. In other words, it belongs to the LLA in terms of $\ln(1+\eta)$, while other terms are further $\alpha_s$-suppressed as they contribute to N$^{k>1}$LA with respect to $\ln(1+\eta)$. Summing the series \ce{eq:C-ser} only for this first term and extending the integration in $y$ up to infinity, which amounts to only adding power-suppressed corrections in $\eta$, one obtains the following resummation formula in the {\it strict LLA in $\ln(1+\eta)$}:
\begin{equation}
    \frac{d\hat{\sigma}^{\text{(HEF, $\ln (1+\eta)$)}}_{\gamma i}}{dz d^2{\bf p}_T} =  \frac{1}{2M^2} \int\limits \frac{d^2{\bf q}_{T1}}{\pi}  {\cal C}_{gi}\left(\frac{1}{1+\eta}, {\bf q}_{T1}^2,\mu_F^2,\mu_R^2\right) \int\limits_{1/z}^{\infty} \frac{dy}{y}\frac{d{\cal H}\big(y,z,{\bf q}_{T1},{\bf p}_T\big)}{dz d^2{\bf p}_T}. \label{eq:sig-HEF-LLA-eta}
\end{equation}

In this approximation, there is no longitudinal integration connecting the resummation part with the HEF coefficient function (``projectile impact-factor'' in the BFKL terminology), which illustrates the connection of our formalism with small-$x$ resummation provided by the CGC/dipole-model framework. The possibility to upgrade the latter framework by restoring the above mentioned longitudinal-momentum integral has been recently discussed~\cite{Boussarie:2021wkn}. We also note that the development of the formalism to perform the NLL computations in the original HEF framework~\cite{Catani:1990xk,Catani:1990eg,Collins:1991ty,Catani:1994sq} of \ce{eq:master-sig-HEF} was recently significantly advanced~\cite{Nefedov:2020ecb,vanHameren:2022mtk} with a clarification of the mechanism of the cancellation of the divergences at NLL. In Section~\ref{sec:results}, we will compare the numerical results of the $\ln(1/\xi)$ and $\ln(1+\eta)$ LLA formalisms matched to the NLO CF computation and will assess the phenomenological relevance of the differences between them for the total cross section of inclusive quarkonium photoproduction.

\subsection{Resummation functions in the {DLA}}

In the framework of CF at leading twist, there are two kinds of large perturbative corrections which enter the cross sections at large $\sqrt{s_{\gamma p}}$. First, there are the corrections enhanced by logarithms of $1+\eta$ and which we have discussed in Section~\ref{sec:HEF-basic} and \ref{sec:strict-LLA}. These lead to the perturbative instability of the quarkonium-production cross section. A second type of large logarithms enters  the DGLAP evolution of PDFs as corrections enhanced by $\ln 1/z$ ($z$ being the parton light-cone momentum fraction) to the DGLAP splitting functions~\cite{Jaroszewicz:1982gr,Catani:1994sq,Altarelli:1999vw} which are functions of $z$. Most of the existing fits of collinear PDFs do not take into account these corrections to the DGLAP splitting functions to all orders in $\alpha_s$ because the evolution of these PDFs is governed by the fixed-order NLO or NNLO DGLAP splitting functions. The resummation of $\ln(1/z)$-enhanced corrections in the PDF evolution has proven to be a complicated task, requiring a non-trivial matching of the BFKL to the DGLAP series~\cite{Altarelli:1999vw} which only relatively recently has led to significant improvements in the quality of PDF fits~\cite{Ball:2017otu, Abdolmaleki:2018jln}. 

In the case of $p_T$-integrated quarkonium-production cross sections, the perturbative instability of the cross section sets in at relatively modest collision energies, see e.g. \cf{fig:TCS_Jpsi_Y1S_NLO}. The fixed-order PDF evolution is still valid at these values of $x$. The PDF uncertainties are thus relatively small and the usage of high-energy improved PDFs~\cite{Ball:2017otu} does not resolve the instability of NLO cross section~\cite{ColpaniSerri:2021bla}. Instead, the corrections leading to the perturbative instability come from the high-energy behaviour of the coefficient function.  Our goal  is thus to take them into account consistently with the fixed-order NLO and NNLO PDF evolutions. 

As we recently discussed~\cite{Lansberg:2021vie}, in order to achieve the goal stated in the previous paragraph, one cannot use the full LLA of HEF. {Instead, one shall use} the resummation functions ${\cal C}_{gg}$ and ${\cal C}_{gq}$ of the HEF formalism in the DLA, which resums terms scaling like $\left(\alpha_s \ln (1/x) \ln(\mu_F^2/{\bf q}_T^2)\right)^n$ to all orders in perturbation theory via the Bl\"umlein-Collins-Ellis formula~\cite{Blumlein:1995eu}:
\begin{equation}
{\cal C}^{\rm (DL)}_{gg}(x,{\bf q}_T^2,\mu^2_F,\mu^2_R)=\frac{\hat{\alpha}_s}{{\bf q}_T^2} \left\{ \begin{matrix}
J_0\left( 2\sqrt{\hat{\alpha}_s \ln\left(\frac{1}{x}\right) \ln \left(\frac{\mu_F^2}{{\bf q}_T^2} \right) } \right) & \text{if }{\bf q}_T^2<\mu_F^2, \\
I_0\left( 2\sqrt{\hat{\alpha}_s \ln\left(\frac{1}{x}\right) \ln \left(\frac{{\bf q}_T^2}{\mu_F^2} \right) } \right) & \text{if }{\bf q}_T^2>\mu_F^2,
\end{matrix} \right. \label{eq:Bluemlein}
\end{equation}
where $\hat{\alpha}_s=\alpha_s(\mu_R)C_A/\pi$, and $J_0$ ($I_0$) are the Bessel functions of the first (second) kind\footnote{Despite an apparent non-smoothness at ${\bf q}_T^2=\mu_F^2$, both functions actually have the same series expansion in $\hat{\alpha}_s \ln \mu_F^2/{\bf q}_T^2$, which is convergent for all ${\bf q}_T^2$.}. Any corrections in the LLA beyond this approximation will be inconsistent either with the $\mu_F$ dependence or with the factorisation scheme on which most of the existing NLO and NNLO PDFs are based, except {the specific PDFs} coming from Refs.~\cite{Ball:2017otu, Abdolmaleki:2018jln} . 

For the case of quark-induced channel, the resummation factor in the LLA (and DLA) has to be modified as:
\begin{equation}
 {\cal C}_{gq}(x,{\bf q}_T^2,\mu^2_F,\mu^2_R)=\frac{C_F}{C_A}\left[  {\cal C}_{gg}(x,{\bf q}_T^2,\mu^2_F,\mu^2_R) - \delta(1-x)\delta({\bf q}_T^2)\right],  \label{eq:C-gq}
\end{equation}
which corresponds to the leading in $k^+$ gluon emission\footnote{Which is described by the ``target impact factor'' in BFKL formalism.} being replaced by the quark one, while in the $t$-channel only gluons still propagate in the leading power (Eikonal) approximation with respect to $x$.

The DLA resummation factors \ce{eq:Bluemlein} and \ce{eq:C-gq} have the following remarkable properties:
\begin{eqnarray}
\int\limits_0^{\mu_F^2}d{\bf q}_T^2\ {\cal C}^{\rm (DL)}_{gg} (x,{\bf q}_T^2,\mu_F^2,\mu_R^2) &=& \delta(1-x), \label{eq:C-gg-norm} \\
\int\limits_0^{\mu_F^2}d{\bf q}_T^2\ {\cal C}^{\rm (DL)}_{gq} (x,{\bf q}_T^2,\mu_F^2,\mu_R^2) &=& 0, \label{eq:C-gq-norm}
\end{eqnarray}
which can be most easily proven using their Mellin-space representations. We will often rely on these properties in the calculations below.

\subsection{Numerical implementation of HEF and cross checks}

To implement numerically the resummation formulae \ce{eq:master-sig-HEF} and \ce{eq:sig-HEF-LLA-eta}, we have to deal with the strong oscillatory behaviour of the resummation function \ce{eq:Bluemlein} at ${\bf q}_T^2\to 0$. Fortunately, the convergence properties of the integrals to be computed can be significantly improved using the properties of \ce{eq:C-gg-norm} and \ce{eq:C-gq-norm}. To this end, we add and then subtract the small-${\bf q}_T$ limit of the HEF coefficient function ${\cal H}$ multiplied by $\theta(\mu_F^2-{\bf q}_{T1}^2)$ in \ce{eq:master-sig-HEF} to get:
\begin{eqnarray}
&&\hspace{-10mm}\frac{d\hat{\sigma}_{\gamma i}^{\text{(HEF, $\ln 1/\xi$)}}}{dz d^2{\bf p}_T} = \frac{1}{2M^2} \int\limits_{\xi_{\min}}^1 \frac{d\xi}{\xi}\  \delta_{ig}\delta(\xi-1) \int\limits_0^{2\pi} \frac{d\phi}{2\pi}\frac{d{\cal H}(\xi(1+\eta),z,{\bf q}_{T1}^2= 0)}{dz d^2{\bf p}_T} + \frac{d\check{\sigma}_{\gamma i}^{\text{(HEF, $\ln 1/\xi$)}}}{dz d^2{\bf p}_T} , \label{eq:sigHEF-LOexpl}\\
&&\hspace{-10mm}\frac{d\check{\sigma}_{\gamma i}^{\text{(HEF, $\ln 1/\xi$)}}}{dz d^2{\bf p}_T}= \frac{1}{2M^2} \int\limits_{\xi_{\min}}^1 \frac{d\xi}{\xi} \int\frac{d^2{\bf q}_{T1}}{\pi} {\cal C}_{gi}^{\text{(DL)}}(\xi,{\bf q}_{T1}^2,\mu_F^2,\mu_R^2) \nonumber \\ &&\hspace{-10mm}\times \left[ \frac{d{\cal H}(\xi(1+\eta),z,{\bf q}_{T1}^2)}{dz d^2{\bf p}_T} - \frac{d{\cal H}(\xi(1+\eta),z,{\bf q}_{T1}^2=0)}{dz d^2{\bf p}_T}\theta(\mu_F^2-{\bf q}_{T1}^2)\right].\label{eq:sigHEF-LOsubtr}
\end{eqnarray}
The first integral over $\xi$ in \ce{eq:sigHEF-LOexpl} is trivially removed due to the $\delta$ function and this term reproduces the known LO CF result, with the LO coefficient function equal to
\begin{equation}
    c_0(\eta, z_{\max})=\frac{1}{2M^2 F_{\rm LO}(1+\eta)} \int\limits_0^{z_{\max}}dz  \frac{d{\cal H}(1+\eta,z,{\bf q}_{T1}^2=0)}{dz}.\label{eq:C0-from-HEF}
\end{equation}
 Due to the on-shell-limit property of the HEF coefficient function of \ce{eq:H-HEF_on-shell},  \ce{eq:C0-from-HEF} exactly reproduces the well-known LO CF scaling function $c_0$~\cite{Kramer:1995nb}. 

The expression in square brackets in the {\it LO-subtracted} HEF result, \ce{eq:sigHEF-LOsubtr}, tends to zero when ${\bf q}_{T1}^2\to 0$. As such, it damps the rapid oscillations of the resummation function, \ce{eq:Bluemlein}, and facilitates the numerical evaluation of the integral \ce{eq:sigHEF-LOsubtr}.

The same procedure should be performed also with the LLA $\ln(1+\eta)$ resummation formula \ce{eq:sig-HEF-LLA-eta}, yielding the result:
\begin{eqnarray}
&&  \frac{d\hat{\sigma}^{\text{(HEF, $\ln (1+\eta)$)}}_{\gamma i}}{dz d^2{\bf p}_T} =  \frac{\delta_{ig} \delta(\eta)}{2M^2} \int\limits_0^{2\pi} \frac{d\phi}{2\pi} \int\limits_{1/z}^{\infty} \frac{dy}{y}\frac{d{\cal H}\big(y,z,{\bf q}^2_{T1}=0,{\bf p}_T\big)}{dz d^2{\bf p}_T} +  \frac{d\check{\sigma}^{\text{(HEF, $\ln (1+\eta)$)}}_{\gamma i}}{dz d^2{\bf p}_T}, \label{eq:sigHEF-LLA-LOexpl}
\end{eqnarray}
where
\begin{eqnarray}  \frac{d\check{\sigma}^{\text{(HEF, $\ln (1+\eta)$)}}_{\gamma i}}{dz d^2{\bf p}_T} &=&  \frac{1}{2M^2} \int\limits \frac{d^2{\bf q}_{T1}}{\pi}  {\cal C}^{\text{(DL)}}_{gi}\left(\frac{1}{1+\eta}, {\bf q}_{T1}^2,\mu_F^2,\mu_R^2\right) \nonumber \\
 &\times& \int\limits_{1/z}^{\infty} \frac{dy}{y} \left[ \frac{d{\cal H}\big(y,z,{\bf q}_{T1},{\bf p}_T\big)}{dz d^2{\bf p}_T} - \frac{d{\cal H}\big(y,z,{\bf q}^2_{T1}=0,{\bf p}_T\big)}{dz d^2{\bf p}_T}\theta(\mu_F^2-{\bf q}_{T1}^2) \right].\label{eq:sigHEF-LLA-LOsubtr}
\end{eqnarray}

The first term in \ce{eq:sigHEF-LLA-LOexpl} is just a crude approximation to the LO CF coefficient function, which corresponds to the LLA $\ln(1+\eta)$. It does not contribute at $\eta\gg 1$ and we will discard it through the matching procedure described in Section~\ref{sec:InEW}. The ${d\check{\sigma}^{\text{(HEF, $\ln (1+\eta)$)}}_{\gamma i}}/{dz d^2{\bf p}_T}$ in \ce{eq:sigHEF-LLA-LOsubtr} is the {\it LO-subtracted HEF result for the strict $\ln(1+\eta)$ resummation} which we will use in the matching procedure below.

Let us emphasise now an important cross check of the resummation formalism that we have employed. The resummation formulae \ce{eq:sigHEF-LOsubtr} or \ce{eq:sigHEF-LLA-LOsubtr}, when expanded up to $O(\alpha_s)$, should reproduce the $\eta\gg 1$ asymptotics of the NLO scaling functions in CF which we have already mentioned in Section~\ref{sec:CF}. We were able to perform this expansion of the scaling functions differential in $z$ and ${\bf p}_T^2$ within the HEF formalism in an analytic form. We have found an excellent agreement with numerical NLO CF results. This agreement constitutes another non-trivial cross check of the HEF formalism at NLO, supplementing other such cross checks at NLO for different processes~\cite{Ellis:1990hw,Catani:1992rn,Ball:2001pq,Lansberg:2021vie} and at NNLO~\cite{Hautmann:2002tu,Marzani:2008uh,Czakon:2013goa,Muselli:2015kba,Marzani:2015oyb,Luo:2019szz}. Moreover, since the DLA for the resummation factors holds up to $O(\alpha_s^3)$, we were also able to numerically compute the $\eta\gg 1$ asymptotics of the $\alpha_s^2\ln(1+\eta)$ term in the NNLO CF coefficient function for $J/\psi$ inclusive photoproduction. We describe all these computations in detail in \ref{AppendixB}.
\section{Matching HEF to NLO CF}
\label{sec:matching+num}

\subsection{Inverse-Error-Weighting matching}
\label{sec:InEW}

In order to combine the NLO CF estimate of $\hat{\sigma}_{\gamma i}(\eta)$ --which is the best source information we have on the behaviour of this quantity for $\eta$ not far from unity-- to the corresponding HEF estimate at $\eta\gg 1$, we use the {\it Inverse-Error-Weighting (InEW)} matching prescription. It has first been introduced in Ref.~\cite{Echevarria:2018qyi} and later improved in our previous paper~\cite{Lansberg:2021vie} where we have provided a self-consistent scheme of estimation of errors entering the InEW weights. The InEW matching is based on a weighted sum of both CF and HEF partonic cross sections such that%\footnote{In the first \textcolor{red}{preprint} version of this paper the matching formula $\hat{\sigma}_{\gamma i}(\eta)=\hat{\sigma}_{\gamma i}^{\text{(CF, LO)}}(\eta)+\left[w_{\gamma i}^{\text{(CF)}}(\eta)\hat{\sigma}_{\gamma i}^{\text{(CF, NLO)}}(\eta) + w_{\gamma i}^{\text{(HEF)}}(\eta)\check{\sigma}_{\gamma i}^{\text{(HEF)}}(\eta)\right]$ had been used. We have verified that the difference of the matched cross sections produced by the \ce{eq:InEW-matching} with respect to the old formula is below 3\%, therefore our numerical results shown in the plots below remain unchanged.}
\begin{equation}
    \hat{\sigma}_{\gamma i}(\eta)=w_{\gamma i}^{\text{(CF)}}(\eta) \, \hat{\sigma}_{\gamma i}^{\text{(CF)}}(\eta) + w_{\gamma i}^{\text{(HEF)}}(\eta)\,\hat{\sigma}_{\gamma i}^{\text{(HEF)}}(\eta),
    \label{eq:InEW-matching}
\end{equation}
with the following prescription for the weight functions:
\begin{equation}
    w_{\gamma i}^{\text{(CF)}}(\eta)=\frac{\left( \Delta \hat{\sigma}_{\gamma i}^{\text{(CF)}}(\eta)\right)^{-2}}{\left( \Delta \hat{\sigma}_{\gamma i}^{\text{(CF)}}(\eta)\right)^{-2}+\left( \Delta \hat{\sigma}_{\gamma i}^{\text{(HEF)}}(\eta)\right)^{-2}}, \ \ w_{\gamma i}^{\text{(HEF)}}(\eta)=1-w_{\gamma i}^{\text{(CF)}}(\eta). \label{eq:InEW-w}
\end{equation}

Correspondingly, the local matching uncertainty in $\eta$ {follows from} the matching procedure and reads:
\begin{equation}
    \Delta \hat{\sigma}_{\gamma i}^{\text{(InEW)}}(\eta) = \left[ \left( \Delta \hat{\sigma}_{\gamma i}^{\text{(CF)}}(\eta) \right)^{-2} + \left( \Delta \hat{\sigma}_{\gamma i}^{\text{(HEF)}}(\eta) \right)^{-2} \right]^{-1/2}.\label{eq:InEW-err}
\end{equation}

{Since the LL($\ln 1/\xi$) resummation explicitly includes the full $\eta$ dependence of the LO CF contribution, as explained in  \ce{eq:sigHEF-LOexpl}, taking into account that $w_{\gamma i}^{\text{(CF)}}(\eta)+w_{\gamma i}^{\text{(HEF)}}(\eta)=1$, one finds that, for this resummation scheme, \ce{eq:InEW-matching} is equivalent to}
\begin{equation}
    \hat{\sigma}_{\gamma i}(\eta)=\hat{\sigma}_{\gamma i}^{\text{(CF, LO)}}(\eta)+\left[w_{\gamma i}^{\text{(CF)}}(\eta)\hat{\sigma}_{\gamma i}^{\text{(CF, NLO)}}(\eta) + w_{\gamma i}^{\text{(HEF)}}(\eta)\check{\sigma}_{\gamma i}^{\text{(HEF)}}(\eta)\right] .\label{eq:InEW-matching_LO-expl}
\end{equation}
{ For the numerical computations below, we use the matching formula \ce{eq:InEW-matching_LO-expl} also in the case of the LL($\ln(1+\eta)$) resummation. As explained by \ce{eq:sigHEF-LLA-LOexpl}, for this resummation scheme, the LO contribution in $\alpha_s$ to the resummed coefficient function is different from the} exact {CF LO }{ one and therefore \ce{eq:InEW-matching_LO-expl} is not strictly equivalent to \ce{eq:InEW-matching}. However, we have verified that, even in this case, the difference of the matched cross sections produced by \ce{eq:InEW-matching_LO-expl} relative to \ce{eq:InEW-matching} is below 3\% because it arises from the power-suppressed high-energy tail of $\hat{\sigma}_{\gamma i}^{\text{(CF, LO)}}(\eta)$.}

An important motivation behind using InEW  is to be able to include, via the error estimates, $\Delta\hat{\sigma}_{\gamma i}^{\text{(CF)}}$ and $\Delta\hat{\sigma}_{\gamma i}^{\text{(HEF)}}$, all the available perturbative information on the effects {\it missing} in each of the contributions in the corresponding limit~\cite{Lansberg:2021vie}. Doing so, we aim to reduce as much as possible the arbitrariness of the matching. The NLO CF contribution is obviously missing any kind of NNLO corrections. At $\eta\gg 1$, the NNLO corrections principally contain the high-energy logarithmic term $\alpha_s^2\ln(1+\eta)$ and  contributions which are constant in $\eta$. We estimate the first from our HEF resummed result and we parameterise the second from the NLO CF result multiplied by $\alpha_s$. We then combine our estimates of these higher-order contributions in quadrature:
\begin{equation}
    \Delta\hat{\sigma}_{\gamma i}^{\text{(CF)}}(\eta)=\sqrt{\left( \alpha_s^2(\mu_R)  C_{\gamma i}^{\text{(LLA)}}(\mu_F) \ln(1+\eta) \right)^2 + \left(\alpha_s(\mu_R)\hat{\sigma}_{\gamma i}^{\text{(CF, NLO)}}(\eta) \right)^2}, \label{eq:delta_CF}
\end{equation}
where the coefficient of the LLA term, $C_{\gamma i}^{\text{(LLA)}}$, is obtained from the expansion of the HEF resummed result \ce{eq:sig-HEF-LLA-eta} for $\hat{\sigma}_{\gamma i}$ up to NNLO:
\begin{equation}
 \begin{split} C_{\gamma i}^{\text{(LLA)}}(\mu_F)=\frac{F_{\text{LO}}}{(2\pi)^2} \left[ c_2^{(\gamma i)}(\infty,z_{\max}) + 2C_A c_1^{(\gamma i)}(\infty,z_{\max}) \ln \left(\frac{M^2}{\mu_F^2}\right)\right. \\ \left. + C_A \bar{c}_1^{(\gamma i)}(\infty,z_{\max}) \ln^2\left(\frac{M^2}{\mu_F^2} \right) \right], 
 \end{split}
\end{equation}
where the asymptotic values of scaling functions $c_1$, $c_2$ and $\bar{c}_1$ are computed numerically, using respectively  \ce{eq:c1-HEFformula}, \ce{eq:c2-HEFformula} and \ce{eq:c1-bar-result}  of \ref{AppendixB}.  
We cannot compute the coefficient that is constant at $\eta\gg 1$ of the NNLO cross section so far as it belongs to NLL HEF. The second term under the square root in \ce{eq:delta_CF} thus provides a generic estimate of higher-order corrections to $\hat{\sigma}_{\gamma i}^{\text{(CF)}}$ which are non-logarithmic in $\eta$. It is simply constructed from $\hat{\sigma}_{\gamma i}^{\text{(CF, NLO)}}$, which is constant at $\eta\gg 1$, multiplied by $\alpha_s$ as an estimate of $\alpha_s^2$ corrections.

On the other hand, our HEF calculation is done only in the LLA. Corrections beyond LLA are thus missing. 
Let us also stress that any HEF computation at any logarithmic accuracy is accurate up to power corrections in $\eta$. It is therefore natural to account for both uncertainties for the missing logarithms and for the missing power corrections in $\eta$. Combining the estimates of uncertainties from these two sources we get: 
\begin{equation}
    \Delta\hat{\sigma}_{\gamma i}^{\text{(HEF)}}(\eta) = \sqrt{\left( \alpha_s(\mu_R)\hat{\sigma}_{\gamma i}^{\text{(CF, NLO)}}(\eta) \right)^2 + \left(C_{\gamma i}^{\rm (HEF)} \eta^{-\alpha_{\gamma i}^{\rm (HEF)}} \right)^2  },
\end{equation}
where the first term under the square root weighs roughly the unknown NLLA corrections, whose perturbative expansion  starts from a constant $O(\alpha_s^2)$ term at $\eta\gg 1$, and the second term stands for the power corrections in $\eta$ missing in the HEF. We compute the exponent $\alpha_{\gamma i}^{\rm (HEF)}>0$ and the normalisation factor $C_{\gamma i}^{\rm (HEF)}$ numerically from the deviation of the known NLO CF ($O(\alpha_s)$) coefficient function from its high-energy limit: $\hat{\sigma}_{\gamma i}^{\text{(CF, NLO)}}(\eta) - \hat{\sigma}_{\gamma i}^{\text{(CF, NLO)}}(\infty)$.

\begin{figure}[hbt!]
    \centering
    \subfloat[]{
       \includegraphics[width=0.6\textwidth]{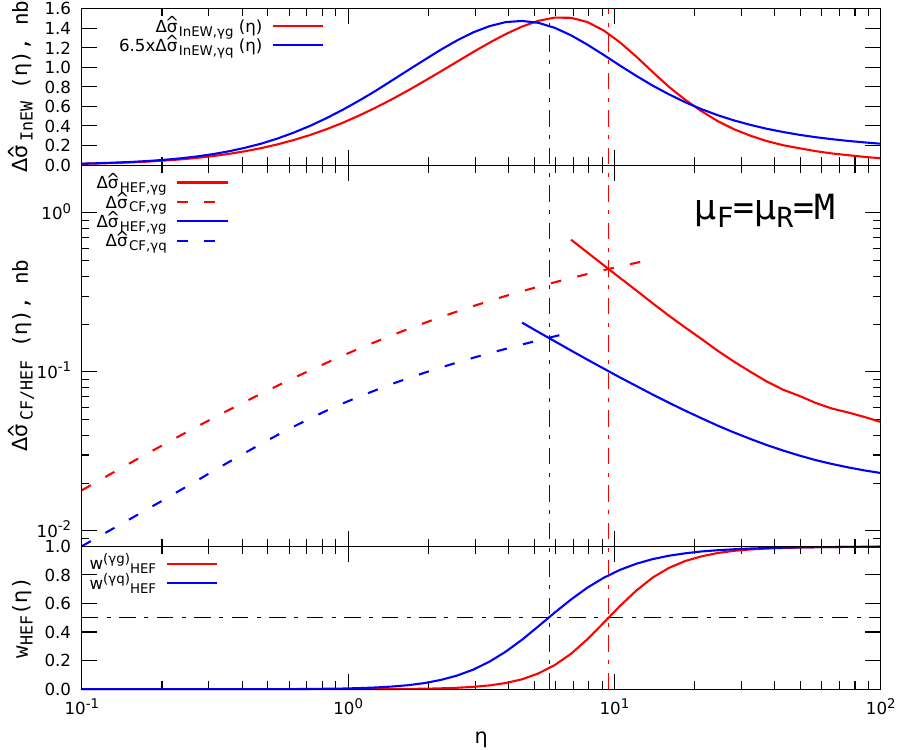}} \\
         \subfloat[]{\includegraphics[width=0.7\textwidth]{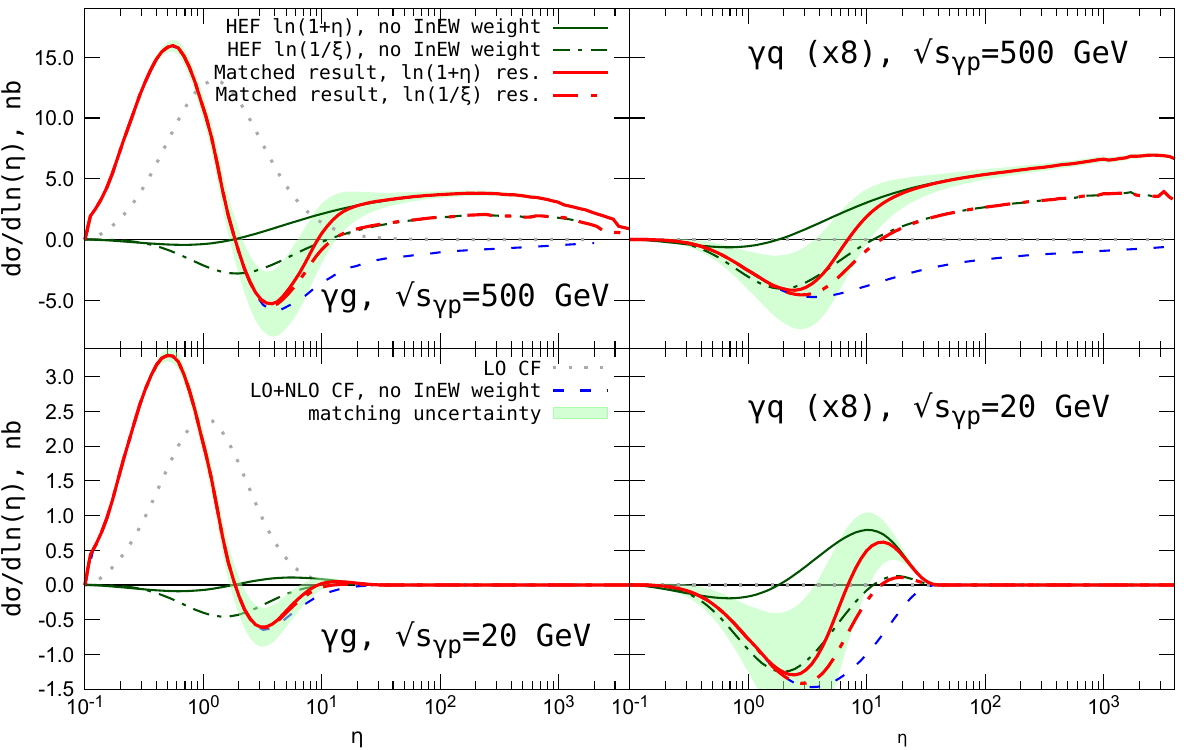}}
    \caption{(a): (top) matching uncertainties, $\Delta\hat{\sigma}_{\gamma i}^{\rm (InEW)}$, (center) 
    Error estimates $\Delta\hat{\sigma}_{\gamma i}^{\rm (CF)}$ and $\Delta\hat{\sigma}_{\gamma i}^{\rm (HEF)}$ and (bottom) resulting InEW weights of HEF contributions, $w_{\gamma i}^{\text{(HEF)}}(\eta)$. (b): $\eta$ integrand of the expression for the total cross section, \ce{eq:TCS-int-eta}, with the LO CF, NLO CF, HEF and matched approximations for $\hat{\sigma}_{\gamma i}$ as well as of the corresponding matching uncertainty for two values of $\sqrt{s_{\gamma p}}$, 500 GeV (top) and 20 GeV (bottom). The plots for $\gamma q$ channel are multiplied by a factor 8 for visibility.}
    \label{fig:InEW-illustration}
\end{figure}

The behaviour of the error estimates obtained as described above together with the resulting InEW weight functions for the HEF contributions is illustrated in the \cf{fig:InEW-illustration}(a). The transition between CF and HEF contributions happens around the point where $\Delta\hat{\sigma}_{\gamma i}^{\text{(CF)}}\simeq \Delta\hat{\sigma}_{\gamma i}^{\text{(HEF)}}$, which occurs at $\eta\simeq 10$ for the $\gamma g$ channel which is dominant at high $\sqrt{s_{\gamma p}}$ and at $\eta\simeq 6$ for the $\gamma q$ channel. The bulk of the matching uncertainty, \ce{eq:InEW-err}, is concentrated in the neighbourhood of this point as can be seen in \cf{fig:InEW-illustration}. The relative magnitude of the CF and HEF contributions to the final result can be seen in the \cf{fig:InEW-illustration}(b). For $\sqrt{s_{\gamma p}}<1$ TeV, the HEF contribution, albeit being very significant, is still smaller than the CF one. As a consequence, for all practically available energies, the matching of HEF contributions to CF ones is necessary and the HEF calculation alone cannot provide a reliable result.

\subsection{Numerical results and theory uncertainties. The dynamical scale-choice.}
\label{sec:results}

For the numerical computation of the cross sections, we have used the multi-threaded version of the \texttt{vegas} Monte-Carlo integration algorithm with cross checks using the \texttt{suave} algorithm as implemented in the \texttt{CUBA} library~\cite{CUBA}. The relative uncertainty of the integration for the total cross sections is below 1\%. To obtain the  numerical results shown in this section, we have used the \texttt{CT18NLO} PDF set~\cite{Hou:2019efy} and the corresponding value of $\alpha_s(M_Z)$. 

The numerical results for the total cross section of inclusive $J/\psi$ photoproduction as the function of the photon-proton collision energy($\sqrt{s_{\gamma p}}$) are shown in the  \cf{fig:TCS_plots_Jpsi_zmax-09_NO-DATA}. We first concentrate on the left panel where the results with the default (conventional) scale-choice, $\mu_F=\mu_R=M$, are shown. The red hatched bands indicate the $\mu_R$ and $\mu_F$ scale dependence about the central-scale result using the same 5-point scale-variation prescription as in \cf{fig:TCS_Jpsi_Y1S_NLO} in  Section~\ref{sec:CF}. The key feature of the matched result in \cf{fig:TCS_plots_Jpsi_zmax-09_NO-DATA} is of course that the scale-variation band does not have the pathological high-energy behaviour of the fixed-order result shown in \cf{fig:TCS_Jpsi_Y1S_NLO}(a).

We also plot in  \cf{fig:TCS_plots_Jpsi_zmax-09_NO-DATA} a separate $\mu_F$-variation band using a green hatched bend. The latter one shows that the $\mu_F$ dependence of the matched cross section at high energy is dramatically reduced in comparison with the $\mu_F$ dependence of the LO result. This comes from the partial cancellation of the $\mu_F$ dependence of the resummation factor of \ce{eq:Bluemlein} and of the PDF. The latter observation illustrates the consistency of the DLA resummation scheme used for the present study with PDF fixed-order evolution. 

In  \cf{fig:TCS_plots_Jpsi_zmax-09_NO-DATA}, we also compare our HEF results, obtained with the central $\mu_R$ and $\mu_F$ scale choices, for both the $\ln(1+\eta)$ (\ce{eq:sigHEF-LLA-LOsubtr}) and the $\ln(1/\xi)$ (\ce{eq:sigHEF-LOsubtr}) resummation matched to NLO CF.  The former is depicted by the solid red lines  and the latter by the dash-dotted red lines. As we have discussed in Section~\ref{sec:strict-LLA}, the result from the $\ln(1/\xi)$ resummation contains some NLL contributions relative to the $\ln(1+\eta)$ resummation. It appears that the difference between both results lies well within the scale-variation band of the $\ln(1+\eta)$ resummation result (see the red hatched band  of \cf{fig:TCS_plots_Jpsi_zmax-09_NO-DATA}). This can be seen as a hint that the NLL HEF corrections are under control in our matching approach. A stronger statement would require matching a complete NLL HEF to (N)NLO CF computations.

\begin{figure}
    \centering
    \begin{tabular}{cc}
      \includegraphics[width=0.47\textwidth]{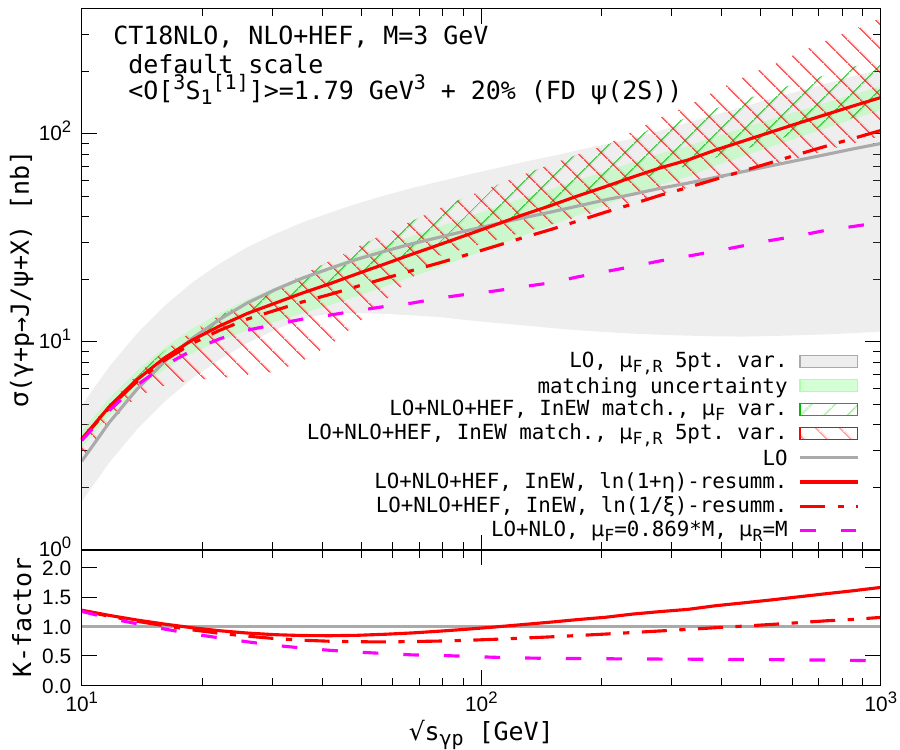}  & \includegraphics[width=0.47\textwidth]{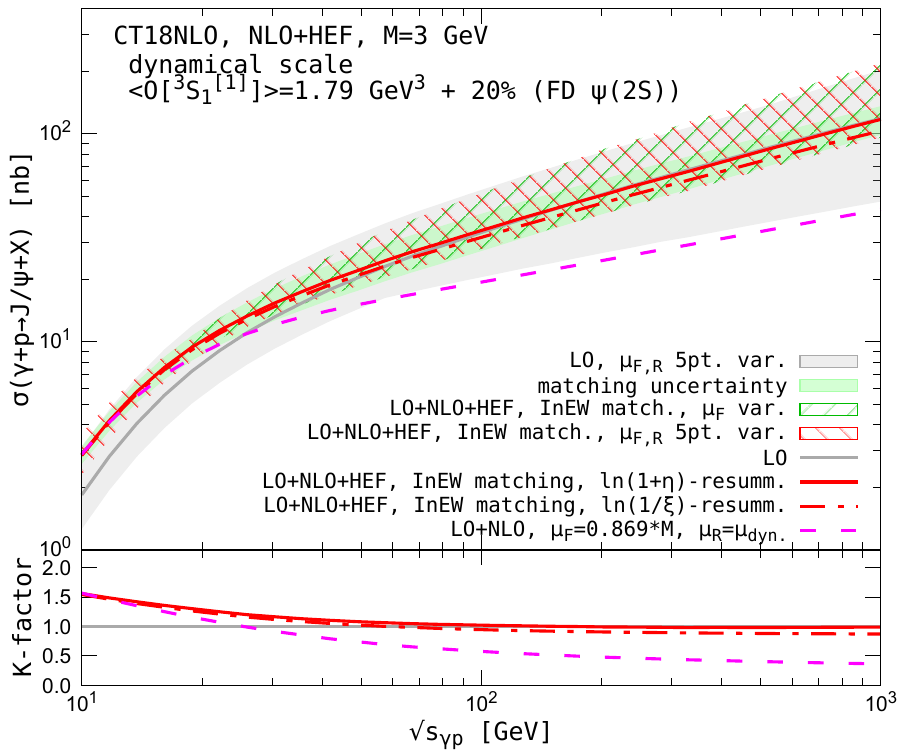} \\
       (a)  & (b)
    \end{tabular}
    \caption{
    Total $J/\psi$-photoproduction cross section for $z<0.9$ obtained via the matching of our HEF-resummed results matched to NLO CF ones using the default scale choice (a) and the dynamical scale choice (b). The red solid line corresponds to the $\ln(1+\eta)$ resummation and the red dash-dotted line to the $\ln(1/\xi)$ resummation. 
    }
    \label{fig:TCS_plots_Jpsi_zmax-09_NO-DATA}
\end{figure}

In  \cf{fig:TCS_plots_Jpsi_zmax-09_NO-DATA}, we also plot the matching uncertainty estimated with the help of \ce{eq:InEW-err}. It turns out to be comparable to the residual $\mu_F$ uncertainty while being significantly larger than the corresponding uncertainty estimated in our previous study~\cite{Lansberg:2021vie} of $\eta_Q$ hadroproduction where it was found to be negligible. This observation points at a stronger sensitivity of the process at hands to the details of the matching procedure due to the complicated non-monotonous shape of the functions involved (see the right panel of \cf{fig:InEW-illustration}), which was less of a problem in the computation for $\eta_Q$ hadroproduction. That being said, the matching uncertainty is still reasonably small and the scale uncertainty remains the main source of theoretical uncertainties. These are expected to be mitigated only by increasing the perturbative accuracy of the computation. 

The eye-catching drawback of the prediction in the left plot of  \cf{fig:TCS_plots_Jpsi_zmax-09_NO-DATA} is the unphysically-looking dip in the red-hatched scale-variation band for $\sqrt{s_{\gamma p}}$ between 20 and 100 GeV. This dip arises from the large (negative) contribution of the loop corrections to $\hat{\sigma}_{ij}^{\text{(CF)}}(\eta)$ at $\eta\simeq 3$ (\cf{fig:InEW-illustration}, right panel) whose $\mu_R$ dependence is not compensated by the running of $\alpha_s$. This becomes clearly problematic when $\mu_R<M$ as it was already discussed in \cite{ColpaniSerri:2021bla}. In this region of relatively small $\eta$, the leading-power approximation of HEF is certainly invalid and can not be invoked to cure this issue. In addition, the region of large $\eta$, where HEF is valid, does not sufficiently contribute such that the $\eta$-integrated hadronic cross section obtained in our matched computation be different enough from the NLO CF result and thus be insensitive to this feature of the loop correction at moderate values of $\sqrt{s_{\gamma p}}$. Perhaps, the inclusion of threshold effects or high-energy resummation at subleading power in $\eta$ could solve this problem for $\mu_R<M$ which is de facto used when varying the scale $\mu_R$ by a factor 2 about the ``conventional'' choice $M$. 

Yet, as discussed in~\cite{ColpaniSerri:2021bla}, the natural scale for the reaction which we discuss, whose Born contribution is a $2 \to 2$ scattering (\ce{proc:ga+gq-cc+X}), is likely not $M$, even for $p_T$-integrated observables. It is rather the invariant mass of the partonic system, $\sqrt{\hat s}$, as the quarkonium is never produced alone. Indeed, in NRQCD, at least one non-soft gluon has to be emitted to photoproduce a heavy-quark pair in the ${}^3S_1^{[1]}$ state.
 
 Therefore, it is natural to choose the average value of  $\sqrt{\hat{s}}$ obtained from the LO CF subprocess, \ce{proc:gamma+g-cc+g}, for the central values for the scales $\mu_R$ and $\mu_F$.  We find that our dynamical scale ranges\footnote{The behaviour of $\left\langle \hat{s}_{\gamma g} \right\rangle$ as a function of $\sqrt{s_{\gamma p}}$ in the LO CF approximation of the CSM is well described by the following parametrisation: $\left\langle \hat{s}_{\gamma g} \right\rangle(\sqrt{s_{\gamma p}}) =M^2+ \left(\kappa_1 L + (\left\langle \hat{s} \right\rangle_{\infty}-M^2) \kappa_2 L^3 \right)/(1+\kappa_2 L^3)$ with $L=\ln \sqrt{s_{\gamma p}}/M$, and $\left\langle \hat{s} \right\rangle_{\infty}=25$ GeV$^2$, $\kappa_1=7$ GeV$^2$ and $\kappa_2=0.03$ for $M=3$ GeV while $\left\langle \hat{s}\right\rangle_{\infty}=250$ GeV$^2$, $\kappa_1=50$ GeV$^2$ and $\kappa_2=0.1$ for $M=9.5$ GeV.  } for the $J/\psi$ case from 3 GeV {at low energies} to 5~GeV for the highest hadronic energies($\sqrt{s_{\gamma p}}$) we will consider and from 9.5 to 16~GeV {respectively} for the $\Upsilon$ case.     

With this dynamical-scale choice,  $\mu_R$ values close to $M/2$ are not used at mid and large $\sqrt{s_{\gamma p}}$. Consequently, the dip in the scale-variation band simply disappears, see the right panel of  \cf{fig:TCS_plots_Jpsi_zmax-09_NO-DATA}. Let us stress that results with both central-scale choices, $M$ vs $\sqrt{\hat{s}_{\gamma g}}$, are compatible within the scale uncertainty. One notable difference is indeed the disappearance of the dip, the second is that the results of $\ln(1+\eta)$ and $\ln(1/\xi)$ resummations  get closer to each other with the dynamical-scale choice. 

We recall that simply increasing the value of the scale in the NLO CF computation does not help to solve the problem of negative cross sections at high energy (\cf{fig:TCS_Jpsi_Y1S_NLO}). On the contrary the values of the NLO CF cross section  become negative for $\mu_F>M$. The scale choice $\hat{\mu}_F$ of \ce{eq:muF-hat-def},  which is optimal from the point of view of the NLO CF computation~\cite{ColpaniSerri:2021bla}, is smaller than $M$ and leads to cross sections which lie at the lower edge of the LO CF scale-uncertainty band. These are clearly below our matching predictions even with the corresponding scale uncertainty, see the dashed purple line in both panels of  \cf{fig:TCS_plots_Jpsi_zmax-09_NO-DATA}.

\begin{figure}[hbt!]
    \centering
    \begin{tabular}{cc}
    \includegraphics[width=0.47\textwidth]{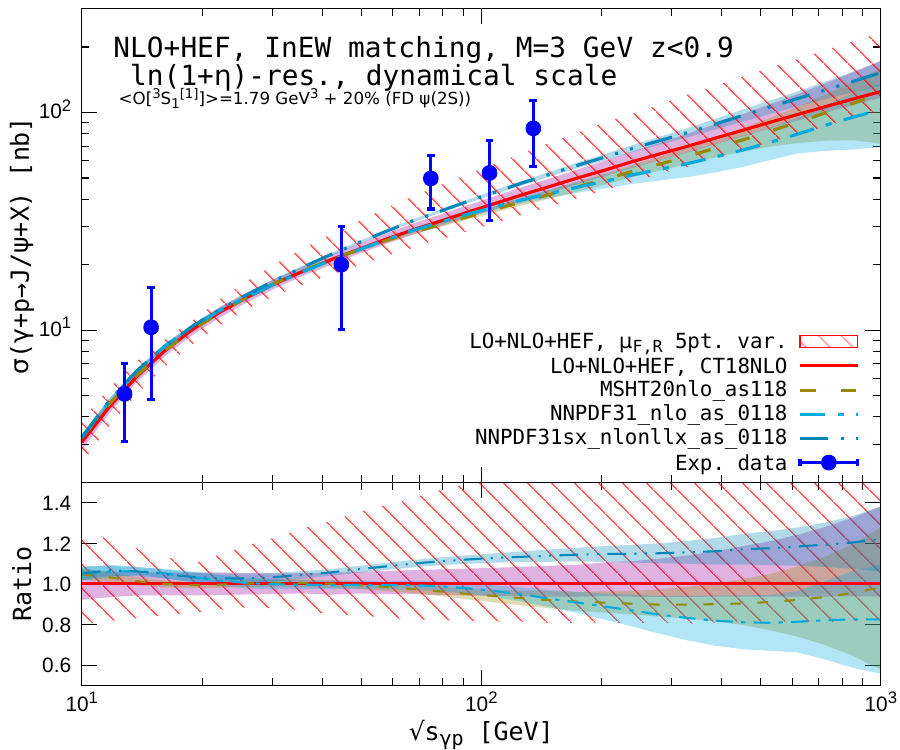} & \includegraphics[width=0.47\textwidth]{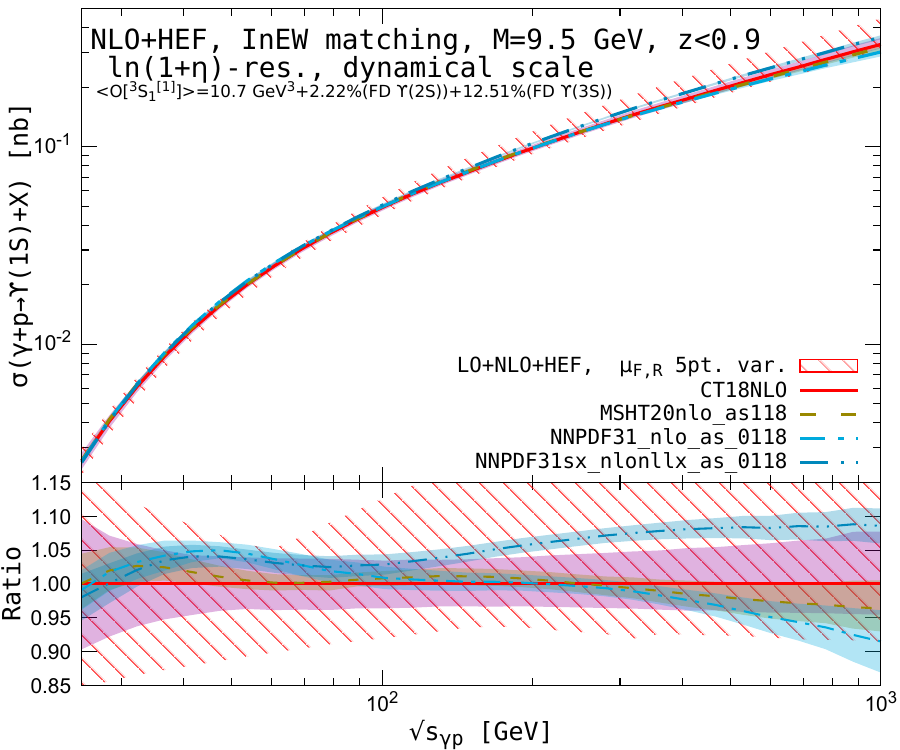} \\
    (a) & (b)
    \end{tabular}
    \caption{Total inclusive photoproduction cross section of $J/\psi$ (a) and $\Upsilon(1S)$ (b) with $z<0.9$ as a function of $\sqrt{s_{\gamma p}}$ in the DLA HEF($\ln(1+\eta)$) matched to NLO CF with our dynamical-scale choice together with their scale variation (hatched band) and PDF uncertainties (solid bands). }
    \label{fig:predictions_Jpsi-Upsilon}
\end{figure}

Having discussed our parameter choices, we are now in a position to present our final matched results. In \cf{fig:predictions_Jpsi-Upsilon}, we show our matched predictions with the dynamical-scale choice for the total inclusive $J/\psi$ (\cf{fig:predictions_Jpsi-Upsilon}(a)) and $\Upsilon(1S)$ (\cf{fig:predictions_Jpsi-Upsilon}(b)) photoproduction cross sections. We focus on the DLA HEF $\ln(1+\eta)$-resummation computation matched to NLO CF as described in Section~\ref{sec:InEW} and use the dynamical-scale prescription of Section~\ref{sec:results}. The scale-variation envelope (see the red-hatched band) in \cf{fig:predictions_Jpsi-Upsilon} is computed with the \texttt{CT18NLO} PDF set.  We have also used the three  PDF sets  which we used in our $\eta_c$-hadroproduction study~\cite{Lansberg:2021vie}, \texttt{MSHT20nlo\_as118}~\cite{Bailey:2020ooq}, \texttt{NNPDF31\_nlo\_as\_0118}~\cite{NNPDF:2017mvq} and \texttt{NNPDF31sx\_ nlonllx\_as\_0118}~\cite{Ball:2017otu} with the central scales.

From the left panel of \cf{fig:predictions_Jpsi-Upsilon}, one can see that our predictions reproduce well the shape of the $\sqrt{s_{\gamma p}}$ dependence and the magnitude of the H1~\cite{H1:1996kyo}, FTPS~\cite{Denby:1983az} and NA14~\cite{NA14:1986mdd} experimental data shown in the plot, unlike the pure NLO CF results, shown in \cf{fig:TCS_Jpsi_Y1S_NLO}. Our improved study clearly shows that the leading-$v$ NRQCD contributions from the ${}^3S_1^{[1]}$  state, or equivalently the CSM, is sufficient to account for the $J/\psi$ data. Even though colour-octet contributions are not needed here, given the large uncertainties of both our computation and the experimental data, a substantial contributions from these colour-octet states, expected from NRQCD NLO fits~\cite{Lansberg:2019adr} despite being NNLO in $v^2$, cannot be excluded. In any case, it will be necessary to consider them through HEF DLA matched to NLO CF as they are likely plagued by the same high-energy perturbative instability~\cite{Mangano:1996kg}. 

It is also worth noting that the PDF uncertainty of our matched results are smaller than their scale uncertainty in the region where experimental data are available. Results from different PDF sets are roughly compatible with each other, which shows that, in this region, the PDFs are reasonably constrained by the small-$x$ DIS data from HERA and the problem of the CF NLO computation (\cf{fig:TCS_Jpsi_Y1S_NLO}) really comes from the poorly controlled high-energy behaviour of the coefficient function $\hat{\sigma}_{\gamma i}(\eta)$. The PDF uncertainty becomes comparable to the scale uncertainty only above $\sqrt{s_{\gamma p}}\sim 1$~TeV. Future experimental data at higher energies to be collected from ultra-peripheral collisions at the LHC in the collider mode~\cite{TalkInclusiveOniumUPC,InclusiveOniumUPC} before possible LHeC or FCC-eh data, as well as data at low energies from EIC~\cite{Accardi:2012qut,OniumEIC} and fixed-target experiments at the LHC~\cite{Hadjidakis:2018ifr,Brodsky:2012vg}, will allow for more precise theory-data comparison. In the meantime, we are hopeful that theoretical studies could be advanced to higher accuracy. 
\section{Conclusions and outlook}
\label{sec:conclusions}

  In the present paper, we have addressed the high-energy perturbative instability of the total cross section of inclusive photoproduction of vector quarkonia. We have assumed colour-singlet dominance  under NRQCD factorisation as what regards the hadronisation of the $Q\bar{Q}$ pair into a quarkonium state. In other words, we have restricted our analysis to the leading-$v^2$ NRQCD contributions; these come from  the ${}^3S_1^{[1]}$  states of the $Q\bar{Q}$ pair.
  
  The partonic cross section of the process, $\hat \sigma_{\gamma i}$, has been obtained by matching the HEF partonic cross section in the DLA which resums, at all order in $\alpha_s$, {a subset of leading-logarithmic} terms scaling like $\alpha_s^n \ln^{n-1}(\hat{s}/M^2)$  in the $\hat{s}\gg M^2$ limit  to the NLO CF partonic cross section. The HEF resummation has been performed within the DLA in order to remain strictly compatible with the DGLAP evolution of the usual collinear PDFs. The matching has been performed using the InEW matching prescription. We have found that the matching is likely not the main source of the theoretical uncertainties of our final results. 

  Our study leads to a solid conclusion that the resummation of high-energy logarithms in the coefficient function of CF solves the perturbative instability of the NLO CF computation at $\sqrt{s_{\gamma p}}\gg M$. It also yields an increase of the cross section up to values compatible with experimental data (within large experimental and theoretical uncertainties). In addition, it significantly reduces the $\mu_F$ dependence of the cross section in this region in comparison to the LO CF prediction. Another important lesson which one should take from the present paper, as well as from our earlier study~\cite{Lansberg:2021vie}, is that, although the HEF contribution to the cross section becomes very significant at high $\sqrt{s_{\gamma p}}\gg M$, the NLO CF contribution from $\sqrt{\hat{s}_{\gamma i}}\gg M$ at $\sqrt{s_{\gamma p}}$ as large as $1$ TeV remain significant. In other words, predictions from HEF or $k_T$-factorisation taken alone~\cite{Lipatov:2004ar,Kniehl:2006sk} would not be sufficient.  

  Our results with the default scale choice $\mu_F\simeq\mu_R\simeq M$ are now well behaved at high $\sqrt{s_{\gamma p}}$ (\cf{fig:TCS_plots_Jpsi_zmax-09_NO-DATA}).
  However, as we discussed in Section~\ref{sec:results}, we consider that using the invariant mass of the partonic scattering as a dynamical scale choice is better  motivated since the CS ${}^3S_1^{[1]}$ vector $S$-wave $Q\bar{Q}$ state is always produced in association with at least one hard gluon. As such, the partonic invariant mass is always larger than the quarkonium mass, $M$. Our final predictions are evaluated using this scale choice (\cf{fig:predictions_Jpsi-Upsilon}) and they agree well with the existing experimental data.
Not only did negative cross sections disappear and did the $\mu_F$ dependence decrease, but also theory now fully agrees with data.

  Our research program of studying quarkonium-production cross sections with matched computations of HEF DLA to NLO CF can be expediently continued in several directions, from resolving the same perturbative instabilities in the $J/\psi$ total {\it inclusive} hadroproduction~\cite{Feng:2015cba,Lansberg:2010cn} or {\it exclusive} photoproduction~\cite{Flett:2021ghh,Flett:2021xsl,Flett:2019pux,Jones:2016ldq,Jones:2015nna,Ivanov:2004vd} cross sections, to reconciling the behaviour of quarkonium $p_T$-distributions at moderate $p_T\lesssim M$ with NRQCD. However, in order to improve the predictive power of the proposed formalism and to reduce its $\mu_R$ dependence, it is desirable to go beyond the DLA on the resummation side. 

\section*{Acknowledgements}
We thank V. Bertone, M.~Fucilla, L. Szymanowski, Y. Yedelkina for useful discussions and Y.~Feng for having shared his FDC results with us.
This project has received funding from the European Union's Horizon 2020 research and innovation programme under grant agreement No.~824093 in order to contribute to the EU Virtual Access {\sc NLOAccess} and the JRA Fixed-Target Experiments at the LHC and a Marie Sk{\l}odowska-Curie action ``RadCor4HEF'' under grant agreement No.~101065263.
This project has also received funding from the Agence Nationale de la Recherche (ANR) via the grant ANR-20-CE31-0015 (``PrecisOnium'') and via the IDEX Paris-Saclay "Investissements d’Avenir" (ANR-11-IDEX-0003-01) through the GLUODYNAMICS project funded by the "P2IO LabEx (ANR-10-LABX-0038)".
This work  was also partly supported by the French CNRS via the IN2P3 project GLUE@NLO, via the Franco-Chinese LIA FCPPL (Quarkonium4AFTER) and via the Franco-Polish EIA (GlueGraph).

\appendix

\section{Derivation of the ${\bf p}_T$-integrated HEF coefficient function\label{AppendixA}}

In order to integrate \ce{eq:H-cal-def} over ${\bf p}_T$, we first split the integrations over the azimuthal angle $\phi$ and ${\bf p}_T^2$. Due to the identity (\ref{eq:qT1.pT-sub}), the $\phi$ dependence can be eliminated from the factor $H$ and, hence, the only $\phi$-dependent factor under the angular integral will be the $\delta$ function:
\begin{equation}
\int\limits_0^{2\pi}d\phi\ \delta\left( \frac{1-z}{z}(M^2(yz-1)-{\bf p}_T^2)-{\bf q}_{T1}^2-{\bf p}_T^2+2|{\bf q}_{T1}||{\bf p}_T|\cos\phi \right) = \frac{2\theta(D)}{\sqrt{D}},\label{eq:B-ang-int}
\end{equation}
where
\[
D=4{\bf p}_T^2{\bf q}_{T1}^2 - \left( {\bf p}_T^2+{\bf q}_{T1}^2 + \frac{1-z}{z}({\bf p}_T^2 - M^2(yz-1)) \right)^2.
\]
The requirement $D>0$ leads to the following upper and lower limits of the ${\bf p}_T^2$ integration:
\[
{\bf p}_{T\pm}^2=\frac{1}{{\bf q}_{T1}^2}\left( z{\bf q}_{T1}^2 \pm \sqrt{D_1} \right)^2,
\] 
with $D_1={\bf q}_{T1}^2(1-z)(M^2(yz-1)-z{\bf q}_{T1}^2)$. From the requirement $D_1>0$, it follows that ${\bf q}_{T1}^2 \leq M^2(y z-1)/z$.

Parametrising ${\bf p}_T^2={\bf p}_{T-}^2 + x({\bf p}_{T+}^2-{\bf p}_{T-}^2)$ with $0\leq x\leq 1$, one finds that the factor $\sqrt{D}$ from \ce{eq:B-ang-int} reduces to $4\sqrt{D_1 x(1-x)}$ while the rest of the dependence of the integrand on $x$ is a rational function. Via a partial-fraction decomposition of the rational dependence on $x$, the integral is expressed as linear combination of   
\begin{equation}
    j_n(a,b)=\int\limits_0^1\frac{dx}{\sqrt{x(1-x)}} \frac{1}{(ax+b)^n} = \frac{\pi}{\sqrt{b(a+b)}}\left\{ \begin{array}{cc}
        1 & \text{ for }n=1  \\
        \frac{a+2b}{2b(a+b)} & \text{ for }n=2 
    \end{array}  \right. , \label{eq:appA:Jn-ints}
\end{equation}
and the ${\bf p}_T$-integrated coefficient function takes the form:
\begin{equation}
\begin{split}
\frac{d{\cal H}(y, z,{\bf q}_{T1}^2)}{dz} =& \frac{\langle{\cal O}[^3S_1^{[1]}]\rangle}{M^3}\frac{64 \pi  \alpha  \alpha_s^2 e_Q^2}{27 yz \left(\tau  (z-1)^2+d_1^2\right)^2 \left(\tau 
   (z-1) (2 \tau  z+z+1)-d_1^2\right)^3} 
   \\
   &\times \Bigl\{ f_1\cdot\Bigl[  j_1 \left(4 d_1 \tau  (z-1),d_1^2-2 d_1 \tau  (z-1)+\tau  (z-1) (\tau 
   (z-2)-1)\right)
   \\
   & + \frac{z}{1-z} j_1 \left(4 d_1 \tau  z,(d_1-\tau  z)^2+\tau \right) \Bigr] 
   \\
   & + f_2\cdot j_2 \left(4 d_1 \tau  (z-1),d_1^2-2 d_1 \tau  (z-1)+\tau  (z-1) (\tau 
   (z-2)-1)\right) 
   \\ 
   & +f_3\cdot j_2 \left(4 d_1 \tau  z,(d_1-\tau  z)^2+\tau\right) \Bigr\},
\end{split}
\end{equation}
where $\tau={\bf q}_{T1}^2/M^2$, $d_1=\sqrt{D_1}/M^2$ and 
\begin{equation}
\begin{split}
&f_1 =2 \tau  (\tau +1) (z-1)^3 z \left(d_1^2-\tau  (z-1) (\tau  z+1)\right)^2
\\
   &\times \left[d_1^4 (\tau +z)-d_1^2 \tau  \left(\tau  \left(z \left(4 z^2-6
   z+5\right)-2\right)+2 \tau ^2 (z-1) z+(z-2) z\right)\right. 
   \\ 
   & \left. +\tau ^2 (z-1) \left(\tau  \left(z
   \left(\tau  (z (z (3 z-8)+8)-2)+(z-2)^2\right)-1\right)-z\right)\right],
   \\
&f_2 = -\tau  (\tau +1) (z-1)^2 \left(d_1^2-\tau  (z-1) (\tau  z+1)\right)^2
\\
&\times   \left[d_1^6 \left(-\left(z^2+2 \tau  (z-1)\right)\right)+d_1^4 \tau  (z-1)
   \left(z^2+2 \tau  (z-1)\right) (4 \tau  z+z+3)\right. 
   \\
& -d_1^2 \tau ^2 (z-1)^2 \left(-6
   \tau +3 \tau ^2 z^4+2 (4 \tau +1) \left(\tau ^2+1\right) z^3 \right. \\
& \left. +(2 \tau  ((7-4 \tau )
   \tau +2)+3) z^2+2 (1-8 \tau ) \tau  z\right)
   \\
& \left.-\tau ^3 (z-1)^3 (2 \tau  z+z+1) \left(2
   \tau  \left(z^2+z+1\right) (z-1)^2-z^2+\tau ^2 z (z (z (z+2)-6)+4)\right)\right],
   \\
&f_3=  \tau  z^2 \left(d_1^2-\tau  (z-1) (\tau  z+1)\right)^2
\\
&\times \left[d_1^6 \left((z-1)^2-\tau  ((z-2) z+3)\right) \right.
\\
& +d_1^4 \tau  (z-1) \left(-(z-1)
   \left(z^2-3\right)+4 \tau ^2 z ((z-2) z+3)-\tau  (z-3) (z (3
   z-2)+3)\right)
   \\
&  -d_1^2 \tau ^2 (z-1)^2 \left(9 \tau +(\tau -1) \tau  (5 \tau +2)
   z^4+2 \tau  (3-5 (\tau -2) \tau ) z^3+ \right.
   \\
& \left. 2 (\tau  (\tau  (7 \tau -12)+3)+1) z^2+12 \tau 
   (2 \tau -1) z-3\right)
   \\
& +\tau ^3 (z-1)^3 (2 \tau  z+z+1) (\tau  (z (\tau  (z (\tau 
   ((z-2) z+2)-(z-6) z-10) \\&+6)+2 (z-3))+3)-1)\Big] .
\end{split}
\end{equation}

If one computes the HEF coefficient function with a minimum ${\bf p}_T^2$ cut such that  ${\bf p}_T^2\geq {\bf p}_{T\min}^2$, one simply has to replace the lower limit of $x$ integration in (\ref{eq:appA:Jn-ints}) by
\begin{equation}
  x_{\min}=\max\left( 0, \frac{{\bf q}_{T1}^2 {\bf p}_{T\min}^2 - (z{\bf q}_{T1}^2 - \sqrt{D_1})^2}{2z{\bf q}_{T1}^2 \sqrt{D_1}} \right).
\end{equation}
Integrals with such a cut also can be expressed in terms of elementary functions.
\section{Derivation of the high-energy asymptotics of the NLO and NNLO CF scaling functions \label{AppendixB} }

In this Appendix, we perform the $\alpha_s$ expansion of the HEF resummation formula \ce{eq:sigHEF-LLA-LOsubtr}. As mentioned before, the $\alpha_s$ expansion of \ce{eq:sigHEF-LOsubtr} is different from the expansion of \ce{eq:sigHEF-LLA-LOsubtr} only by the $\text{N}^{k \geq 1}\text{LLA}$ terms which are outside the scope of the present study. To this end, we rewrite $\theta(\mu_F^2-{\bf q}_{T1}^2)=\theta(M^2-{\bf q}_{T1}^2)+\theta({\bf q}_{T1}^2-M^2) \theta(\mu_F^2-{\bf q}_{T1}^2)$, assuming that $M<\mu_F$, in \ce{eq:sigHEF-LLA-LOsubtr} and use the series expansions in ${\alpha}_s$ of
\begin{eqnarray}
{\cal C}^{\rm (DL)}_{gg}(x,{\bf q}_T^2,\mu_F^2,\mu_R^2)&=&\frac{\hat{\alpha}_s}{{\bf q}_T^2} \left[ 1 +\hat{\alpha}_s \ln \left(\frac{1}{x}\right) \ln\left(\frac{{\bf q}_T^2}{\mu_F^2}\right) + O(\alpha_s^2) \right], \\
\int\limits_{M^2}^{\mu_F^2}d{\bf q}_T^2\  {\cal C}^{\rm (DL)}_{gg}(x,{\bf q}_T^2,\mu_F^2,\mu_R^2) &=& -\hat{\alpha}_s \ln\left(\frac{M^2}{\mu_F^2}\right) - \frac{\hat{\alpha}_s^2}{2}\ln^2\left(\frac{M^2}{\mu_F^2} \right) \ln \left(\frac{1}{x} \right) + O(\alpha_s^3),
\end{eqnarray}
to obtain
\begin{equation}
\begin{split}
\frac{d\check{\sigma}_{\gamma i}^{\text{(HEF, $\ln(1+\eta)$)}}}{dz d^2{\bf p}_T}=& \frac{F_{\rm LO}}{\pi M^2} \Bigg\{ \left(\frac{\alpha_s(\mu_R)}{2\pi}\right) \left[ \frac{d c_1^{(\gamma i)}(\infty,z,\rho)}{dz d\rho} + \frac{d\bar{c}^{(\gamma i)}_1(\infty,z,\rho)}{dz d\rho} \ln\frac{M^2}{\mu_F^2} \right]
\\
&+ \left(\frac{\alpha_s(\mu_R)}{2\pi}\right)^2 \ln(1+\eta) \left[  \frac{d c_2^{(\gamma i)}(\infty,z,\rho)}{dz d\rho} +  2C_A\frac{d c_1^{(\gamma i)}(\infty,z,\rho)}{dz d\rho} \ln \left(\frac{M^2}{\mu_F^2} \right) \right.
\\
&+ \left.   C_A\frac{d \bar{c}_1^{(\gamma i)}(\infty,z,\rho)}{dz d\rho} \ln^2 \left(\frac{M^2}{\mu_F^2} \right)  \right] + O(\alpha_s^3) \Biggr\},
\end{split}
\label{eq:sig-HEF-as2}
\end{equation}
where $\rho={\bf p}_T^2/M^2$. For the asymptotics of the scaling functions in \ce{eq:sig-HEF-as2}, we obtain the following explicit formulae in terms of the HEF coefficient function \ce{eq:H_gamma+R-cc+g}:
\begin{align}
\begin{split}
&\hspace{-6mm}\frac{d\bar{c}^{(\gamma i)}_1(\infty,z,\rho)}{dz d\rho} = \frac{z(1-z)C_i}{8\pi M^2 F_{\rm LO}} \int\limits_0^{2\pi} \frac{d\phi}{2\pi} \left( \lim\limits_{{\bf q}_{T1}^2\to 0} \frac{H(\bar{s},\bar{t},\bar{u},{\bf q}_{T1}\cdot {\bf p}_T,{\bf q}_{T1}^2)}{(1-z+\rho)^2}\right),
\label{eq:c1-bar-HEFformula}
\end{split}
\\
\begin{split}
&\hspace{-6mm}\frac{dc^{(\gamma i)}_1(\infty,z,\rho)}{dz d\rho} = \frac{z(1-z)C_i}{8\pi M^2 F_{\rm LO}} \int \frac{d^2{\bf q}_{T1}}{\pi{\bf q}_{T1}^2}
\\
&\hspace{-6mm}\times \Biggl[ \frac{M^4 H(\bar{s},\bar{t},\bar{u},{\bf q}_{T1}\cdot {\bf p}_T,{\bf q}_{T1}^2)}{\left[(M^2+{\bf p}_T^2)(1-z) + ({\bf q}_{T1}-{\bf p}_T)^2 z\right]^2} - \theta(M^2-{\bf q}_{T1}^2)\left( \lim\limits_{{\bf q}_{T1}^2\to 0} \frac{H(\bar{s},\bar{t},\bar{u},{\bf q}_{T1}\cdot {\bf p}_T,{\bf q}_{T1}^2)}{(1-z+\rho)^2}\right)  \Biggr],
\label{eq:c1-HEFformula}
\end{split}
\\
\begin{split}
&\hspace{-6mm}\frac{dc^{(\gamma i)}_2(\infty,z,\rho)}{dz d\rho} = \frac{z(1-z)C_iC_A}{4\pi M^2 F_{\rm LO}} \int \frac{d^2{\bf q}_{T1}}{\pi{\bf q}_{T1}^2} \ln\frac{{\bf q}_{T1}^2}{M^2}
\\
&\hspace{-6mm}\times \Biggl[ \frac{M^4 H(\bar{s},\bar{t},\bar{u},{\bf q}_{T1}\cdot {\bf p}_T,{\bf q}_{T1}^2)}{\left[(M^2+{\bf p}_T^2)(1-z) + ({\bf q}_{T1}-{\bf p}_T)^2 z\right]^2} - \theta(M^2-{\bf q}_{T1}^2) \left(\lim\limits_{{\bf q}_{T1}^2\to 0} \frac{H(\bar{s},\bar{t},\bar{u},{\bf q}_{T1}\cdot {\bf p}_T,{\bf q}_{T1}^2)}{(1-z+\rho)^2} \right) \Biggr],
\label{eq:c2-HEFformula}
\end{split}
\end{align}
where $C_i=\delta_{ig}C_A + \delta_{iq}C_F$ and the Mandelstam variables are given by \ce{eq:stu-sub} with $y=[(M^2+{\bf p}_T^2)(1-z)+({\bf q}_{T1}-{\bf p}_T)^2 z]/[M^2 z (1-z)]$. 

\ce{eq:c1-bar-HEFformula} only involves a simple averaging over the azimuthal angle and can be evaluated to (for $C_A=N_c=3$):
\begin{equation}
\frac{d\bar{c}^{(\gamma g)}_1(\infty,z,\rho)}{dz d\rho} = \frac{16\pi^2 (1-z) z \left(\rho^2 \left(z^2-z+1\right)^2+\rho \left(z^2-2 z+2\right)
   (z-1)^2+(z-1)^4\right)}{(\rho+1)^2 \left(\rho+(z-1)^2\right)^2 (\rho-z+1)^2}.\label{eq:c1-bar-result}
\end{equation}

The ${\bf q}_{T1}$ integrals in \ce{eq:c1-HEFformula} and \ce{eq:c2-HEFformula} are finite in two dimensions thanks to the cancellation of $1/{\bf q}_{T1}^2$ singularity between both terms in the square brackets. As such, these integrals can be easily evaluated numerically. Moreover, one notices that the integrand of \ce{eq:c1-HEFformula} is a rational function of ${\bf q}_{T1}^2$ and ${\bf q}_{T1}\cdot {\bf p}_T$, which suggests the application of standard loop-integral techniques, such as Integration-By-Parts (IBP) Reduction~\cite{Chetyrkin:1981qh}. We use these techniques below in this Appendix to obtain closed-form analytic result of the integrals of \ce{eq:c1-HEFformula}.

In order to be able to split the integration of the first and second terms in \ce{eq:c1-HEFformula} we go to $2-2\epsilon$ dimensions, which will regularise the collinear divergences when the terms are separated. The ${\bf q}_{T1}^2$ integral in the second term can be easily evaluated  and \ce{eq:c1-HEFformula} turns into
\begin{eqnarray}
\frac{dc^{(\gamma i)}_1(\infty,z,\rho)}{dz d\rho} &=& \frac{z(1-z)C_i}{8\pi M^2 F_{\rm LO}}\left[ F_1-F_2 \right], \label{eq:AppB:c1-HEFformula} \\
F_1&=&\int \frac{d^2{\bf q}_{T1}}{\pi{\bf q}_{T1}^2} \frac{M^4 H(\bar{s},\bar{t},\bar{u},{\bf q}_{T1}\cdot {\bf p}_T,{\bf q}_{T1}^2)}{\left[(M^2+{\bf p}_T^2)(1-z) + ({\bf q}_{T1}-{\bf p}_T)^2 z\right]^2} , \label{eq:AppB:F1} \\
F_2&=&-\frac{(M^2)^{-\epsilon}}{\epsilon} \int\frac{d\Omega_{2-2\epsilon}}{2\pi} \left( \lim\limits_{{\bf q}_{T1}^2\to 0} \frac{H(\bar{s},\bar{t},\bar{u},{\bf q}_{T1}\cdot {\bf p}_T,{\bf q}_{T1}^2)}{(1-z+\rho)^2}\right), \label{eq:AppB:F2}
\end{eqnarray}
where $\bar{s},\bar{t},\bar{u}$ are given by \ce{eq:stu-sub} with $y=[(M^2+{\bf p}_T^2)(1-z)+({\bf q}_{T1}-{\bf p}_T)^2 z]/[M^2 z (1-z)]$ and $d\Omega_{2-2\epsilon}$ is an element of the solid angle describing the direction of the vector ${\bf q}_{T1}$ in $2-2\epsilon$ dimensions.

It turns out not to be necessary to recompute the HEF coefficient function \ce{eq:H-def} in $4-2\epsilon$ dimensions for the computation of the integrals \ce{eq:AppB:F1} and \ce{eq:AppB:F2} because the original finite integral (\ref{eq:c1-HEFformula}) is two-dimensional. We have redone the calculation with the $4-2\epsilon$-dimensional version of $H$ and have obtained the same results.

One has however to be careful with the evaluation of \ce{eq:AppB:F2} because the averaging over the directions of the vector ${\bf q}_{T1}$, which is done in this term after taking the limit ${\bf q}_{T1}^2\to 0$, has to be done in $2-2\epsilon$ dimensions to stay consistent with the evaluation of \ce{eq:AppB:F1} within dimensional regularisation. The source of the problem lies in the angular integrations of the type:
\[
  \int\frac{d\Omega_{2-2\epsilon}}{2\pi} ({\bf p}_T\cdot {\bf q}_{T1})^2 =  {\bf p}_T^2 {\bf q}_{T1}^2 \frac{\pi^{-1/2-\epsilon}}{\Gamma(1/2-\epsilon)} \int\limits_0^\pi d\phi \sin^{-2\epsilon}\phi\cdot \cos^2\phi = \frac{{\bf p}_T^2 {\bf q}_{T1}^2}{2(1-\epsilon)} \frac{\Omega_{2-2\epsilon}}{2\pi}, 
\]
with $\Omega_{2-2\epsilon}=2\pi^{1-\epsilon}/\Gamma(1-\epsilon)$ being the solid angle in $2-2\epsilon$ dimensions. The $\epsilon$ dependence coming from these terms leads to finite terms in $F_2$,  scaling like $\epsilon^0$, where $\epsilon$  has been
cancelled.

The dependence on ${\bf q}_{T1}$ in the integrand of \ce{eq:AppB:F1} is rational and one finds the following three denominators in it:
\begin{eqnarray*}
{\cal D}_1&=&(2-z){\bf q}_{T1}^2  - 2({\bf p}_T\cdot{\bf q}_{T1}) +{\bf p}_T^2 + M^2(1-z)  , \\
{\cal D}_2&=& z^2 {\bf q}_{T1}^2  - 2z({\bf p}_T\cdot {\bf q}_{T1}) +{\bf p}_T^2 + M^2(1-z)^2, \\
{\cal D}_3&=& {\bf q}_{T1}^2,
\end{eqnarray*}
which are linearly dependent, since we have only two linearly-independent scalar products ${\bf q}_{T1}^2$ and ${\bf p}_T\cdot{\bf q}_{T1}$. The appearance of linearly-dependent denominators is a common feature of quarkonium-related calculations. Due to this, one has to perform a partial-fractioning decomposition of the integrand in \ce{eq:AppB:F1}. We can then split the integrand into three parts, depending on the combinations $({\cal D}_1,{\cal D}_2)$, $({\cal D}_1,{\cal D}_3)$ and $({\cal D}_2,{\cal D}_3)$. In each of these integral families, the scalar products in the numerator can be uniquely expressed in terms of linear combinations of the denominators. As such, the resulting integals have positive or negative powers of ${\cal D}_i$ and these can then be reduced using IBP reduction codes, such as \texttt{LiteRed}~\cite{LiteRed1,LiteRed2}. The resulting master integrals are then evaluated using Feynman parameters in terms of Gaussian hypergeometric functions ${}_2F_1$ which can be expanded in $\epsilon$ using the \texttt{HypExp}~\cite{Huber:2005yg} package. We have also extensively used various features of the \texttt{FeynCalc} framework~\cite{FeynCalc,Shtabovenko:2016sxi,Shtabovenko:2020gxv} on all stages of this computation.

With the procedure described above, one obtains the following result for the $c_1$ coefficient differential with respect to $z$ and $\rho={\bf p}_T^2/M^2$ with $C_A=N_c=3$,
\begin{equation}
\begin{split}
&\frac{dc^{(\gamma g)}_1(\infty,z,\rho)}{dz d\rho} = 8\pi^2 \Biggl\{ c^{\rm (R)}_1 (z,\rho)
\\
&+ c^{(1)}_1 (z,\rho) \ln \left[ \frac{z^2 (1-z)^2}{(\rho + (1-z)^2)^2} \right] + c^{(2)}_1 (z,\rho) \ln \left[ \frac{(\rho + 1-z)^2}{(1-z)(\rho + 2-z)} \right]
\\
&+ \frac{c^{(3)}_1 (z,\rho)}{\sqrt{(1+\rho)((2-3z)^2 + (2-z)^2\rho)}}
\\
&\times \ln \left[\frac{\rho (2-z)-(3-2 z)
   z+2 -\sqrt{(\rho +1) \left(\rho  (z-2)^2+(2-3 z)^2\right)} }{\rho (2-z)-(3-2 z)z+2 +\sqrt{(\rho +1) \left(\rho  (z-2)^2+(2-3 z)^2\right)} }\right] \Biggr\}, 
\end{split}
\label{eq:c1-result}
\end{equation} 
and $c^{(\gamma q)}_1(\infty,z,\rho)=(C_F/C_A)c^{(\gamma g)}_1(\infty,z,\rho)$, while the rational part of the result is
%{}
\begin{eqnarray*}
&&\hspace{-8mm}c^{\rm (R)}_1 (z,\rho) = -2z\left\{-16 (\rho +1)^6 (\rho  (\rho +2)-1)+8 (\rho  (\rho +4)+9) z^{14} \right. \\
&&\hspace{-8mm} -4 \left(\rho 
   \left(6 \rho ^2+46 \rho +135\right)+195\right) z^{13}   +2 (\rho  (\rho  (5 \rho  (3 \rho
   +44)+942)+1660)+2011) z^{12} \\
&&\hspace{-8mm} -(\rho  (\rho  (\rho  (\rho  (20 \rho
   +551)+3762)+8908)+11638)+13273) z^{11} \\
&&\hspace{-8mm} +(\rho  (\rho  (\rho  (\rho  (\rho  (7 \rho
   +361)+4074)+15298)+23699)+28573)+31476) z^{10} \\
&&\hspace{-8mm} -(\rho  (\rho  (\rho  (\rho  (\rho  (\rho  (\rho
   +112)+2289)+14320)+34205)+40070)+55265)+56458) z^9 \\
&&\hspace{-8mm} +(\rho  (\rho  (\rho  (\rho  (2 \rho  (\rho 
   (5 \rho +292)+3394)+26973)+42734)+46706)+87812)+78121) z^8 \\
%\end{eqnarray*}
%\begin{eqnarray*}
&&\hspace{-8mm} +(\rho  (\rho  (\rho  (\rho  (\rho 
   (\rho  ((\rho -32) \rho -1383)-10257)-24354) \\ &&-20442)-40831)-112901)-83721) z^7 \\
&&\hspace{-8mm} +(\rho  (\rho 
   (\rho  (\rho  (\rho  (\rho  ((25-7 \rho ) \rho +1401)+4754)-5507) \\ &&-24167)+30395)+113588)+69198)
   z^6 \\
&&\hspace{-8mm} +2 (\rho +1) (\rho  (\rho  (\rho  (\rho  (\rho  (\rho  (9 \rho
   +35)+396)+5184) \\ && +17041)+10711)-21550)-21826) z^5 \\
&&\hspace{-8mm} -(\rho +1)^2 (\rho  (\rho  (\rho  (\rho  (\rho 
   (24 \rho +355)+4025)+17018)+21830)-7645)-20631) z^4 \\
&&\hspace{-8mm} +4 (\rho +1)^3 (\rho  (\rho  (\rho  (\rho  (8
   \rho +181)+1176)+2368)+396)-1769) z^3 \\
&&\hspace{-8mm} -4 (\rho +1)^4 (\rho  (\rho  (\rho  (13 \rho
   +179)+579)+337)-416) z^2\\
&&\hspace{-8mm} \left. +16 (\rho +1)^5 (\rho  (\rho  (3 \rho +19)+21)-15) z \right\} \bigg/ \bigg[ (\rho +1)^2   \left(\rho +(z-1)^2\right)^2 (\rho -2 z+1)^2 \\
&& (\rho -z+1)^2 (\rho -z+2) \left(\rho  (z-2)^2+(2-3 z)^2\right)^2 \bigg] ,
\end{eqnarray*}
and the coefficients in front of the logarithmic terms read
\begin{eqnarray*}
&&\hspace{-8mm} c_1^{(1)}(z,\rho) = \frac{-z^3}{(\rho +1)^2 \left(\rho +(z-1)^2\right)^2 (\rho
   -2 z+1)^4} \\
&&\hspace{-8mm}   \times \left\{5 (\rho +1)^4+4 (2 \rho +1) z^6-(\rho +1) (23 \rho +31) z^5+(\rho +1) (\rho  (12
   \rho +77)+89) z^4 \right. \\ 
&&\hspace{-8mm}   \left. -2 (\rho +1) (\rho +3) (\rho  (\rho +18)+21) z^3+2 (\rho +1)^2 (\rho  (3 \rho
   +32)+47) z^2-(\rho +1)^3 (11 \rho +35) z\right\} , \\
\end{eqnarray*}
\begin{eqnarray*}
&&\hspace{-8mm}c_1^{(2)}(z,\rho) = \frac{z}{(\rho +1)^2 (\rho -2 z+1)^4 (\rho -z+1)^2} \\
&&\hspace{-8mm} \times \left\{2 (\rho +1)^4-4 (2 \rho +1) z^6+(7 \rho  (\rho +2)-9) z^5+(\rho  ((5-2 \rho ) \rho
   +48)+57) z^4 \right. \\
&&\hspace{-8mm} \left. +(\rho +1) (\rho  ((\rho -13) \rho -69)-87) z^3+(\rho +1)^2 (\rho  (5 \rho +36)+59)
   z^2-6 (\rho +1)^3 (\rho +3) z\right\}, \\
\end{eqnarray*}
\begin{eqnarray*}
&&\hspace{-8mm} c_1^{(3)}(z,\rho) =  \frac{z^3}{(\rho +1)^2 (\rho -2 z+1)^4 \left(\rho 
   (z-2)^2+(2-3 z)^2\right)^2} \\
&&\hspace{-8mm} \times\left\{-16 (\rho -10) (\rho +1)^5+64 (\rho  (\rho +4)+9) z^9-32 (\rho  (\rho  (3 \rho
   +23)+64)+96) z^8\right. \\
&&\hspace{-8mm} +4 (\rho  (\rho  (\rho  (10 \rho +119)+245)+461)+1093) z^7 \\
&&\hspace{-8mm}  +(\rho  (\rho  (\rho 
   (618-\rho  (7 \rho +135))+7730)+13133)+3461) z^6 \\
&&\hspace{-8mm} +(\rho  (\rho  (\rho  (\rho  (23 \rho
   -397)-8578)-34546)-45573)-18289) z^5 \\
&&\hspace{-8mm} -2 (\rho +1) (\rho  (\rho  (\rho  ((\rho -7) \rho
   -1392)-10488)-21697)-12625) z^4 \\
&&\hspace{-8mm} +2 (\rho +1)^2 (\rho  (\rho  (\rho  (7 \rho
   -99)-2827)-10497)-9200) z^3 \\
&&\hspace{-8mm} \left. -4 (\rho +1)^3 (\rho  (\rho  (9 \rho -112)-1269)-1912) z^2+8 (\rho
   +1)^4 (5 (\rho -11) \rho -214) z\right\}.
\end{eqnarray*}

The coefficients $c_1^{(1)}$, $c_1^{(2)}$, $c_1^{(3)}$  contain the denominator $(\rho -2z +1)^{-4}$ which is singular at the physical point $\rho = 2z-1$, while $c_1^{\rm (R)}$ is proportional to $(\rho -2z +1)^{-2}$. Individual terms may be divergent at this point. However, combining all such terms in the complete expression of $dc_1/dz d\rho$ given in \ce{eq:c1-result}, the divergences cancel against each other and the limit at this point is actually finite,
\begin{eqnarray*}
&& \lim\limits_{\rho\to 2z-1} \frac{dc_1^{(\gamma g)}(\infty,z,\rho)}{dz d\rho} = \frac{\pi^2}{z^8
   (z+1)} \\
&& \left\{ z^3 \left[(z (z (z (z (12 z (2 z-9)+191)-52)-339)+513)-121) \log (1-z) \right. \right. \\
&& -8 (z-1) (z+1) (z (z (2 z (2
   z-5)+13)-8)+2) \log (z) \\
&& \left. +(z (z (z (z (4 z (2 z+7)-119)+68)+251)-449)+105) \log (z+1)\right] \\
&& +2 (z (z (z
   (z (z (z (4 (z-2) z+85)-243)+267)+40)-333)+245)-55) z \\
&& \left.  +2 (z (298 z-245)+55) \tanh ^{-1}(z) \right\},
\end{eqnarray*}
where $\tanh ^{-1}(z) = \ln\left[ (1+z)/(1-z) \right]/2$ is the hyperbolic arc-tangent. 

\begin{figure}
    \centering
    \includegraphics[width=0.5\textwidth]{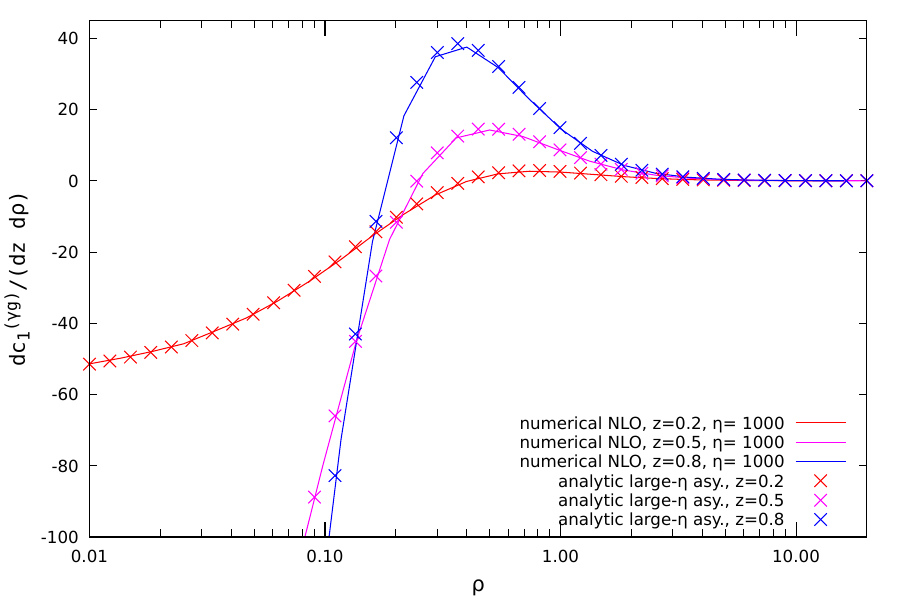}
    \caption{The numerical comparison of the asymptotic result (\ref{eq:c1-result}) with the scaling function computed by our numerical NLO code.}
    \label{fig:c1-diff-num}
\end{figure}

The numerical comparison of our result (\ref{eq:c1-result})  with our calculation of $dc_1^{(\gamma g)}/(dz d\rho)$ using dipole subtraction which was already mentioned in Section~\ref{sec:CF}, is shown in \cf{fig:c1-diff-num} for $\eta=1000$. As one can see, the asymptotic result is in a good agreement with numerical data. We also provide in~\ct{tab:c2} and in \cf{fig:plot_c1_c1-bar_c2} the high-energy asymptotic numerical values of scaling functions, obtained via direct numerical evaluation of \ce{eq:c1-bar-HEFformula}, \ce{eq:c1-HEFformula} and \ce{eq:c2-HEFformula} with a relative accuracy about $10^{-3}$ using the regular algorithm \texttt{cuhre} implemented in the \texttt{CUBA} library~\cite{CUBA}.

\begin{table}[]
    \centering
    \begin{tabular}{|c|c|c|c|c|c|c|c|c|c|}
    \hline $z$             &  0.1   & 0.2   & 0.3   & 0.4   & 0.5   & 0.6   & 0.7   & 0.8   & 0.9 \\
    \hline $dc_2/dz$       &  50.09 & 99.81 & 147.6 & 192.3 & 233.7 & 272.7 & 316.1 & 393.4 & 633.2 \\ 
    \hline $dc_1/dz$       &-0.7195 &-1.394 &-2.063 &-2.820 &-3.832 &-5.573 &-9.554 &-20.65 &-56.55 \\
    \hline $d\bar{c}_1/dz$ & 5.267  & 10.55 & 15.90 & 21.39 & 27.20 & 33.74 & 42.02 & 54.57 & 78.93 \\
    \hline
    \end{tabular}
    \caption{Numerical results for the high-energy asymptotics of the $dc_2/dz$, $dc_1/dz$ and $d\bar{c}_1/dz$ scaling functions}
    \label{tab:c2}
\end{table}

\ce{eq:c2-HEFformula} provides predictions for the non-trivial NNLO scaling function $c_2$ in the high-energy limit in the LLA. In principle, it is possible to derive the closed-form analytic result for the high-energy asymptotics of the NNLO scaling function from it. However, such an analytic result would be too cumbersome to present and due to Gram-determinant singularities it may be more challenging to evaluate it numerically rather than the integral (\ref{eq:c2-HEFformula}) itself. Therefore, we limit ourselves to present here the numerical results for the $z$-differential but $p_T$-integrated scaling function $c_2$ in \cf{fig:plot_c1_c1-bar_c2} and in \ct{tab:c2}. 

\begin{figure}
    \centering
    \includegraphics[width=0.45\textwidth]{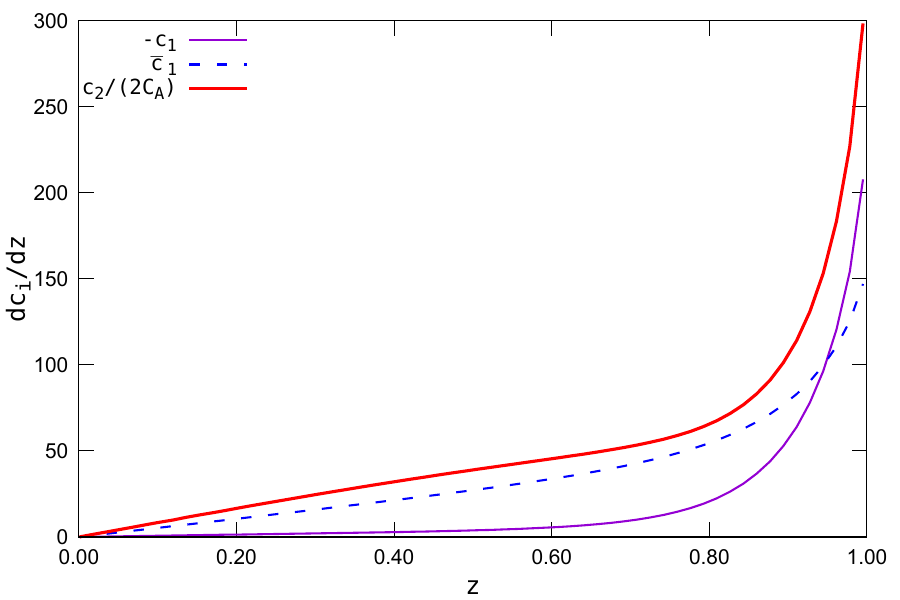}
    \caption{High-energy asymptotics of $z$-differential scaling functions: $dc_1/dz$ (solid line), $d\bar{c}_1/dz$ (dashed line) and $dc_2/dz$ (thick solid line, divided by $2C_A$).}
    \label{fig:plot_c1_c1-bar_c2}
\end{figure}

It is instructive to look at several limits of the obtained results. The low-${\bf p}_T^2$ and high-${\bf p}_T^2$ asymptotics of $\bar{c}_1$ are,
\begin{eqnarray}
\rho\to 0 &:& \frac{d\bar{c}_1(\infty,z,\rho)}{dz d\rho} = \frac{16\pi^2 z}{1-z}+O(\rho), \\
\rho\to \infty &:& \frac{d\bar{c}_1(\infty,z,\rho)}{dz d\rho} = \frac{16\pi^2 z(1-z)}{\rho^4} \left(1-z(1-z)\right)^2 + O(\rho^{-5}), \label{eq:c1-bar-high-pT}
\end{eqnarray}
while for the coefficient $c_1$ we have,
\begin{align}
\begin{split}
\rho\to 0 : \qquad \frac{dc_1(\infty,z,\rho)}{dz d\rho} =& \frac{8\pi^2 z}{(1-2 z)^4}
\\
& \hspace{-2cm}\times \Biggl\{ (64 z^7-256 z^6+116 z^5+653 z^4-1213 z^3+898 z^2-308 z+40)
\\ 
&\hspace{-2cm}\times \frac{z^2}{\left| 2-3 z\right|
   ^3} \ln\left[\frac{-\left| 2-3 z\right| +2 z^2-3 z+2}{\left| 2-3 z\right| +2 z^2-3 z+2}\right]
   \\
&\hspace{-2cm} + \left(4 z^2-15 z+5\right) z^2 \ln\left[ \frac{(z-1)^2}{z^2} \right]
\\
&\hspace{-2cm} + \frac{2\left(2 z^4-3 z^3+7 z^2-5 z+2\right) (1-2 z)^4}{(2-3 z)^2 (z-1)}  
\\
&\hspace{-2cm} \left.  -\frac{\left(4 z^5+13 z^4-44 z^3+43 z^2-16
   z+2\right) }{z-1} \ln\left[ \frac{(1-z)^2}{z^2-3 z+2} \right] \right\} + O(\rho),
   \label{eq:c1-low-pT}
   \end{split}
   \\
   \begin{split}
\rho\to \infty : \qquad \frac{dc_1(\infty,z,\rho)}{dz d\rho} =& \frac{16\pi^2 (1-z) z}{\rho ^3
   (z-2)^2} 
   \\
& \times \left\{ z^4-2 z^3-z^2 \ln \left[\frac{(1-z) z^2}{\rho  (2-z)^2}\right]-4 z+4\right\} + O(\rho^{-4}),
\label{eq:c1-high-pT}
\end{split}
\end{align} 
i.e. the coefficient $c_1$ drops like $1/{\bf p}_T^6$, while $\bar{c}_1\sim 1/{\bf p}_T^8$ at high-${\bf p}_T$ and high partonic energy. 

Another interesting limit is $z\to 1$ where both the ($\rho$-differential) $c_0$ and $\bar{c}_1$ tend to zero, while $c_1$ tends to a non-zero limit,
\begin{equation}
    \lim\limits_{z\to 1}\frac{dc_1(\infty,z,\rho)}{dz d\rho} =\frac{16\pi^2 ((\rho -2) \rho  ((\rho -2) \rho +2)+1)}{(\rho -1)^4 \rho  (\rho +1)^3},\label{eq:c1_z-1_limit}
\end{equation}
such that the vanishing of the LO $Q\bar{Q}\left[{}^3S_1^{[1]}\right]$ photoproduction cross section at $z\to 1$ is violated already at NLO and may actually receive $\ln(1-z)$ corrections at higher orders. 

\bibliography{mybibfile}

\end{document}